\documentclass[submission, creativecommons, withnoncommercial]{eptcs}
\providecommand{\event}{ICE 2020}

\usepackage[utf8]{inputenc}
\usepackage{xargs}
\usepackage{xspace}
\usepackage{url}
\usepackage{wrapfig}
\usepackage{breakurl}             % Not needed if you use pdflatex only.
\usepackage{underscore}           % Only needed if you use pdflatex.

\usepackage[nomargin,multiuser,inline,draft]{fixme}
\fxusetheme{color}
\usepackage{multicol}               % Table multirow/multicolumn support
\usepackage{subcaption}             % Support for subfigures

\usepackage{comment}

\usepackage{dot2texi}

\usepackage[usenames,dvipsnames,svgnames,table]{xcolor} % must be loaded first
\usepackage{tikzsymbols}
\usepackage{tikz}                   % Graphs
\usetikzlibrary{shadows,arrows,shapes,automata,positioning,decorations.markings}
\tikzset{
  component/.style={
      draw,
      fill = white,
      minimum width = 1.5cm,
      minimum height = .5cm,
      drop shadow
    }
}
\tikzset{
  file/.style={
      thin,
      fill = blue!5,
      font = \tt\scriptsize,
      text width = .8cm,
      minimum width = 1.0cm,
      minimum height = .5cm,
      drop shadow
    }
}

\tikzset{
  dataflow/.style={
      thick,
      draw, ->, >=latex,
      dashed,
      DarkOliveGreen
    }
}

\tikzset{
  pipeline/.style={
      thick,
      draw, ->, >=latex,
      double,
      red
    }
}

\usepackage{amsthm,amsmath,amssymb,amsbsy}
\usepackage{mathtools} % \xRightArrow and stuff - load late to avoid conflicts
\usepackage{thmtools, thm-restate, thm-autoref}  % Advanced theorem handling
\newtheorem{definition}{Definition}
\newtheorem{theorem}{Theorem}

\newtheorem{example}{Example}

\usepackage[capitalise]{cleveref}
\usepackage{enumitem}              % More enumeration styles
\newlist{steps}{enumerate}{1} % also creates a counter called stepsenumi'
\setlist[steps]{label*=\arabic*., ref=(\arabic*)}
\crefalias{stepsi}{step}

\DeclareMathSizes{10}{10}{6}{6}

\usepackage{todonotes}

\usepackage{xargs}
\newif\ifemi
\emifalse
\newcommand{\tnxbehapi}[1][partially]{
  Research {#1} supported by the EU
  H2020 RISE programme under the Marie Skłodowska-Curie grant
  agreement No 778233
}

\newcommand{\tnxitmatters}[1][Work partially funded by]{
  #1 MIUR project PRIN 2017FTXR7S \emph{IT MATTERS}
  (Methods and Tools for Trustworthy Smart Systems)
}

\newcommand{\tnxtrustfull}[1][partially supported by the]{
  {#1} TrustFull project,
  funded by the Swedish Foundation for Strategic Research
}

\DeclareGraphicsExtensions{%
    .png,.PNG,%
    .pdf,.PDF,%
    .jpg,.mps,.jpeg,.jbig2,.jb2,.JPG,.JPEG,.JBIG2,.JB2}

\usepackage[UKenglish]{babel}
  
\usepackage{fixme}
\fxusetheme{color}
\FXRegisterAuthor{eM}{aeM}{\color{orange} {\underline{eM}}}

\usepackage[normalem]{ulem} % underline command breaks over line ends

\usepackage{xifthen}        % for conditional commands
\newcommand{\ifempty}[3]{%
  \ifthenelse{\isempty{#1}}{#2}{#3}%
}

\def\colorFun{\color{orange}}

\newcommand{\mkfun}[4][\colorFun]{
  \newcommand{#2}[1][#4]{
    {\colorFun #1}\textsf{#3}
    \ifempty{##1}{}{
      ({##1})}
  }
}

\newcommand{\mkuop}[4][\colorFun]{
  \newcommand{#2}[1][#4]{
    {#1\textsf{#3}}
    \ifempty{##1}{}{
      \, {##1}}
  }
}

\newcommand{\hidden}[1]{}
\newcommand{\hide}[1]{}

\newcommand{\cf}[2]{
  \fontsize{#1}{#1}{\selectfont{#2}}
}
\ifemi
\usepackage{showlabels}

\newcommand{\emi}[2]{
  \marginpar{\fcolorbox{red}{shadecolor}{\cf{#1}{{#2}}}}
}
\newcommand{\emic}[2]{\par
  \fcolorbox{red}{shadecolor}{\parbox{\linewidth}{ 
      \color{gray}
      \begin{description}
      \item[{\color{blue} #2}]{\sf #1}
      \end{description}}}
}
\else
\newcommand{\emi}[2]{}
\newcommand{\emic}[2]{}{}
\fi

\newcommand{\multiset}[1]{\left\{\!\!\mid {#1} \mid\!\!\right\}}
\newcommand{\multicup}{\Cup}
\newcommand{\multidiff}{-}
\newcommand{\sst}{\;\big|\;}
\newcommand{\qst}{\;\colon\;} %such that

\newcommand{\conf}[1]{\ensuremath{\langle {#1} \rangle}}

\newcommand{\bnfdef}{\ ::=\ }
\newcommand{\bnfmid}{\;\ \big|\ \;}

\newcommand{\qqand}{\qquad\text{and}\qquad}

\newcommand{\nat}{\mathbb{N}}
\newcommand{\upd}[3]{{#1}[{#2} \mapsto {#3}]}

{\bfseries}{\rmfamily}
{\bfseries}{\rmfamily}

\newcommand{\quo}[1]{\lq\lq {#1}\rq\rq}
\def\finex{{\unskip\nobreak\hfil
\penalty50\hskip1em\null\nobreak\hfil$\diamond$
\parfillskip=0pt\finalhyphendemerits=0\endgraf}}

\definecolor{shadecolor}{rgb}{1,0.99,0.9}
\definecolor{bg}{rgb}{0.95,0.95,0.95}

\newcommandx{\testsys}[2][1=\aM,2=\atestcase,usedefault=@]{#1 \otimes  #2}

\def\colorSbj{\color{black}}
\newcommand{\subject}[1][\ae]{\colorSbj{\textsf{sbj}(#1)}}
\newcommandx{\aQfinal}[1][1=,usedefault=@]{
  {\ifempty{#1}{F}{F_{#1}}}
}
\newcommand{\gsubs}[2]{^{#1} / _{#2}}
\newcommandx{\gsubst}[3][1=\aM,2=q,3=q',usedefault=@]{
  \left \{\gsubs{#3}{#2} \right \}#1
}
\newcommandx{\gsubsts}[5][1=\aM,2=q,3=q',4=q,5=q',usedefault=@]{
  \left \{\gsubs{#3}{#2}, \gsubs{#5}{#4} \right \}#1
}

\newcommand{\lact}{\lset_{\text{act}}}
\newcommand{\atestcase}[1][T]{#1}

\newcommand{\Qsucc}[1][Q]{\underline{#1}}

\usepackage{listings}
\usepackage{mfirstuc}
\usepackage{xstring}

\newcommand{\cfsmtr}[1][]{\xrightarrow{#1}}
\newcommand{\aTrs}{\cfsmtr}
\newcommandx{\aCFSM}[3][1=\aQ,2=\aQzero,3=\aTrs]{(#1,#2,#3)}
\newcommand{\cstr}[1]{{\xRightarrow{\raisebox{-.3ex}[0pt][0pt]{$\scriptstyle #1$} }}}

\newif\ifcp
\cpfalse
\newcommand{\gname}[1][i]{\ifcp{\colorNode{\scriptstyle\textsf{#1}}}\else{}\fi}

\newif\ifguard
\guardfalse
\newcommand{\aguard}{\ifguard{\colorGuard \phi}\else{}\fi}

\def\colorGuard{\color{cyan}}
\def\colorPtp{\color{blue}}

\def\colorOp{\color{OliveGreen}}
\def\colorNode{\color{LightCoral}}
\def\colorR{\color{OliveGreen}}
\def\colorE{\color{orange}}

\def\colorMsg{\color{BrickRed}}

\newcommand{\fillcolor}{orange!5}

\newcommand{\keywordcol}{orange}%{ForestGreen!100!blue!80}

\newcommand{\mkkeyword}[1]{\textcolor{\keywordcol}{\textsf{#1}}}

\newcommand{\msg}[1][m]{\mathtt{\colorMsg{#1}}}
\newcommand{\msgset}{\mathcal{M}}
\newcommand{\chset}{\mathcal{C}}

\newcommand{\lset}{\mathcal{L}}

\newcommand{\ptp}[1][A]{\ensuremath{\mathsf{\colorPtp{\capitalisewords{#1}}}}}

\newcommand{\p}{\ptp}
\newcommand{\q}{{\ptp[B]}}

\newcommandx{\ggcommon}[3][1=\ptp,2={\aH},3={\aH'},usedefault=@]{f_{#1}}
\newcommandx{\opair}[2][1={\ae},2={\ae'},usedefault=@]{\conf{{#1},{#2}}}
\newcommandx{\hopair}[2][1={\aE},2={\aE'},usedefault=@]{\llparenthesis\, {#1},{#2}\, \rrparenthesis}
\newcommandx{\wf}[2][1={\aG},2={\aG'},usedefault=@]{wf({#1}, {#2})}
\newcommandx{\wb}[2][1={\aG},2={\aG'},usedefault=@]{wb({#1}, {#2})}
\newcommandx{\ws}[2][1={\aG},2={\aG'},usedefault=@]{ws({#1}, {#2})}
\newcommandx{\widx}[2][1={\aW},2={i},usedefault=@]{{#1}[{#2}]}
\newcommandx{\outop}[2][1=\gname,2={}]{{\colorOp{!}}^{{#1}{#2}}}
\newcommandx{\inop}[2][1=\gname,2={}]{{\colorOp{?}}^{{#1}{#2}}}
\newcommandx{\aout}[5][1=a,2=b,3={},4=m,5={},usedefault=@]{
  \achan[#1][#2] \outop[{#3}] {\msg[#4]} {#5}
}
\newcommandx{\ain}[5][1={\p},2={\q},3=\gname,4=m,5={},usedefault=@]{
  \achan[#1][#2] \inop[{#3}] {\msg[{#4}]}{#5}
}
\newcommandx{\adep}[1][1={}]{
  \conf{ \aout[@][@][@][@][{#1}], \ain[@][@][@][@][{#1}]}
}

\newcommandx{\hproj}[2][1=\aH, 2=\ptp, usedefault=@]{
  \ifempty{#1}{}{{#1}}\ifempty{#2}{}{{^{\scriptscriptstyle @{#2}}}}
}
\newcommandx{\eproj}[2][1=\aE,2=A, usedefault=@]{
  {{#1}}\ifempty{#2}{}{{^{\scriptscriptstyle @{{\ptp[#2]}}}}}
}

\newcommand{\apom}{r}
\newcommand{\emptypom}{\epsilon}

\newcommand{\projpom}[2]{{#1}\!\!\downharpoonright_{#2}}

\newcommand{\aR}[1][R]{{\colorR{#1}}}

\newcommand{\aConf}{s}

\newcommandx{\detM}[1][1=\aCM,usedefault=@]{\Delta({#1})}
\mkfun{\wellformed}{\textit{WF}}{}
\mkfun{\termstates}{\mbox{$\Omega$}}{}
\mkfun{\subgraphs}{\mbox{$\mathbb{T}$}}{}
\newcommand{\atree}{\tau}
\mkfun{\mrun}{$\Pi$}{}
\newcommand{\arun}{\pi}

\tikzset{
  cfsm/.style={
	 node distance=2.2cm and 1cm,
	 scale=.85,
	 transform shape,
	 smooth,
	 every state/.style = {cnode},
	 every edge/.style = {carrow}
  }
}

\tikzset{
  cnode/.style={
    shape=circle,
    minimum size = 0mm,
    inner sep = 1pt,
    font=\tiny,
    draw
  },
  carrow/.style={
    ->,
    shorten >=1pt,
    >=stealth',
    auto,
    font=\scriptsize,
    draw,
    sloped
  }
}

\newcommandx{\cm}[2][1=\ptp, 2=\aM]{{#2}_{#1}}
\newcommandx{\achan}[2][1=A,2=B,usedefault=@]{{\ptp[#1]}{\,}{\ptp[#2]}}
\newcommand{\ptpset}{\mathcal{\colorPtp{P}}}

\newcommand{\oact}{\outop[]}
\newcommand{\iact}{\inop[]}

\newcommand{\csconf}[2]{\conf{\vec{#1} \ ; \ \vec{#2}}}

\newcommand{\trans}[2][{}]{\,\xrightarrow{#2}_{#1}\,}

\newcommandx{\acfsmout}[3][1=A,2=B,3=m,usedefault=@]{\achan[{#1}][{#2}] \oact {\msg[{#3}]}}
\newcommandx{\acfsmin}[3][1=A,2=B,3=m,usedefault=@]{\achan[{#1}][{#2}] \iact {\msg[{#3}]}}
\newcommandx{\fsaout}[2][1={\p},2={},usedefault=@]{
  \ptp[#1] \ \outop[]\ \msg[{#2}]
}
\newcommandx{\fsain}[2][1={\p},2={},usedefault=@]{
  \ptp[#1] \ \inop[]\ \msg[{#2}]
}

\makeatletter
\newcommand{\linenumfontsize}{\@setfontsize{\linenumfontsize}{3pt}{3pt}}
\makeatother
\lstdefinelanguage{sys}{
commentstyle=\color{Gray},
morecomment=[s]{[}{]},
keywords=[0]{system,of,do,end},
keywordstyle=\color{orange}\bfseries,
}

\lstdefinelanguage{sgg}{
commentstyle=\color{Gray},
morecomment=[l]{..},
morecomment=[s]{[}{]},
keywords=[0]{repeat,branch,sel},
keywordstyle=\color{orange}\bfseries,
morekeywords=[1]{*,\+,|,->},
literate={->}{$\colorOp \xrightarrow$}1 {|}{$\gparop$}1 {;}{$\gseqop$}1 {+}{$\gchoop$}1 {\{}{{\textcolor{NavyBlue}{\{}}}1 {\}}{{\textcolor{NavyBlue}{\}}}}1
}

\newcommand{\aG}{\mathsf{G}}

\newcommand{\gseqop}{{\colorOp ;}\,}
\newcommand{\gparop}{{\colorOp \ |\ }}
\newcommand{\gchoop}{{\colorOp \ +\ }}
\newcommand{\grecop}{{\colorOp *}}
\newcommand{\grecopp}{{\colorOp{@}}}
\newcommandx{\nmerge}[2][1={i},2={},usedefault=@]{
\ifempty{#2}{
\ifempty{#1}{\mu}{\gname[-{#1}]}
}{-{#2}}
}

\newcommand{\grepkw}{\mkkeyword{repeat}}
\newcommand{\gblk}[1][\aG]{\textcolor{NavyBlue}{\{} #1 \textcolor{NavyBlue}{\}}}

\mkfun{\esbj}{sbj}{\ae}

\makeatletter%
\@ifclassloaded{exam-paper}%
{}%
{\makeatletter%
  \@ifclassloaded{test}%
  {}%
  \makeatother%
}
\makeatother%

\newcommandx{\gnode}[2][1=i,2=\gint,usedefault=@]{
  \ifcp{
    \ifempty{#1}{#2}{\gname[#1].\big({#2}\big)}
  }
  \else
  {#2}
  \fi
}

\newcommand{\gempty}{\mathtt{(o)}}
\newcommandx{\gint}[4][1=i,2=A,3=m,4=B,usedefault=@]{
  \ptp[#2] {\colorOp \xrightarrow{\msg[{#3}]}} \ptp[#4]
}
\newcommandx{\gout}[4][1=\gname,2=\ptp,3=m,4={\ptp[C]},usedefault=@]{
  \achan[{#2}][{#4}] {\colorOp {\colorOp{!}}} {\msg[{#3}]}
}
\newcommandx{\gin}[4][1=\gname,2=\ptp,3=m,4={\ptp[C]},usedefault=@]{
  \achan[{#2}][{#4}] {\colorOp {\colorOp{?}}} {\msg[{#3}]}
}
\newcommandx{\gseq}[3][1=\gname,2={\aG},3={\aG'},usedefault=@]{
  \def\ggraph{{#2} \gseqop {#3}}
  \ggraph
}
\newcommand{\ginfix}[4]{
  \def\ggraph{{#2} {#4} {#3}}
  \gnode[#1][\ggraph]
}
\newcommandx{\gpar}[3][1=i,2={\aG},3={\aG'},usedefault=@]{
  \ginfix{#1}{#2}{#3}{\gparop}
}
\newcommandx{\gcho}[3][1=i,2={\aG},3={\aG'},usedefault=@]{
  \ginfix{#1}{#2}{#3}{\gchoop}
}
\newcommandx{\gchov}[3][1=\gname,2={\aG},3={\aG'},usedefault=@]{
  \def\ggraph{\left(
    \begin{array}l
      \ifempty{#1}{{#2} \\ \gchoop \\ {#3}}{\!\!{#2} \\ \gchoop \\ {#3}}
    \end{array}\right)}
  \ifcp\gnode[{$#1$}][\ggraph] \else \ggraph \fi
}
\newcommandx{\grec}[3][1=i,2={\aG},3={\p},usedefault=@]{
  \def\ggraph{\grecop {#2} \ifempty{#3}{}{\grecopp {#3}}}
  \ifempty{#1}{\ggraph}{\gname[{$#1$}][\ggraph]}
}

\newcommand{\getcentroid}[2]{
  \coordinate (tmpgatecoord) at (0,0);
  \foreach \n [count=\i] in {#1}{
      \path (\n);
      \coordinate (tmpgatecoord) at ($(tmpgatecoord) + (\n)$);
      \coordinate (#2) at ($1/\i*(tmpgatecoord)$);
    }
}

\tikzset{
  hgsem/.style={
      draw,
      node distance=2cm and 1cm,
      transform shape,
      smooth,
      every node/.style = {font=\sffamily\bfseries}
    }
}

\tikzset{
  hgstyle/.style={
      src color={#1},
      tgt color={#1},
      centroid color={#1},
      centroid label={#1},
      centroid name={#1},
      centroid radius={#1},
      centroid ratio={#1},
      xoffset={#1},
      yoffset={#1},
      xsrcoffset={#1},
      ysrcoffset={#1},
      xtgtoffset={#1},
      ytgtoffset={#1},
      font={#1},
      centroid angle={#1},
      centroid tolerance={#1}
    },
  src color/.store in = \hgsrccol,
  tgt color/.store in = \hgtgtcol,
  centroid color/.store in =\hgfillcolor,
  centroid label/.store in =\hglabel,
  centroid name/.store in =\hgname,
  centroid radius/.store in = \hgradius,
  centroid ratio/.store in = \hgratio,
  xoffset/.store in =\hgxoffset,
  yoffset/.store in =\hgyoffset,
  xsrcoffset/.store in =\hgxsrcoffset,
  ysrcoffset/.store in =\hgysrcoffset,
  xtgtoffset/.store in =\hgxtgtoffset,
  ytgtoffset/.store in =\hgytgtoffset,
  centroid angle/.store in =\hgangle,
  centroid tolerance/.store in =\hgtolerance,
  src color = black,
  tgt color = black,
  centroid color = orange!40,
  centroid label={},
  centroid name={dummycentroid},
  centroid radius = .7pt,
  centroid ratio = .35,
  xoffset = 0,
  yoffset = 0,
  xsrcoffset = 0,
  ysrcoffset = 0,
  xtgtoffset = 0,
  ytgtoffset = 0,
  font=\sffamily\scriptsize,
  centroid angle=0,
  centroid tolerance=10pt
}

\newcommandx{\mkhg}[5][1={},4={},5={},usedefault=@]{
  \begingroup
  \tikzset{#1}
  \StrCount{#2,}{,}[\l] % from package xxstring
  \StrCount{#3,}{,}[\m] % from package xxstring
  \ifthenelse{\l = 1 \AND \m = 1}{
    \ifempty{#4}{
      \ifempty{#5}{
        \path[hgsem, ->, >=stealth', shorten >=1pt] (#2) -- (#3);
      }{
        \path[hgsem, ->, >=stealth', shorten >=1pt] (#2) #5 (#3);
      }
    }{
      \ifempty{#5}{
        \path[hgsem, ->, >=stealth', shorten >=1pt, #4] (#2) -- (#3);
      }{
        \path[hgsem, ->, >=stealth', shorten >=1pt, #4] (#2) #5 (#3);
      }
    }
  }{
    \coordinate (srcoffset) at (\hgxsrcoffset,\hgysrcoffset);
    \coordinate (tgtoffset) at (\hgxtgtoffset,\hgytgtoffset);
    \getcentroid{#2}{srccentroid};
    \getcentroid{#3}{tgtcentroid};
    \node[label={left:\hglabel}] (\hgname) at ($(srccentroid)!{1-\hgratio}!\hgangle:(tgtcentroid) + (\hgxoffset,\hgyoffset)$) {};
    \pgfgetlastxy \xc \yc;
    \pgfmathtruncatemacro{\xcontrol}{\xc};
    \pgfmathtruncatemacro{\ycontrol}{\yc};
    \foreach \n in {#2}{
        \path (\n);
        \pgfgetlastxy \xntmp \yntmp;
        \pgfmathtruncatemacro{\xn}{\xntmp};
        \pgfmathtruncatemacro{\yn}{\yntmp};
        \pgfmathsetmacro\xtmpdiff{abs(\xn - \xcontrol + \hgxsrcoffset)};
        \pgfmathsetmacro\ytmpdiff{abs(\yn - \ycontrol + \hgytgtoffset)};
        \ifdim \xtmpdiff pt > \hgtolerance
          \ifempty{#4}{
            \path[hgsem, \hgsrccol] (\n) .. controls ($(srccentroid.center) + (srcoffset)$) .. (\hgname.center);
          }{
            \path[hgsem, \hgsrccol] (\n) .. controls ($(srccentroid.center) + (srcoffset)$) .. (\hgname.center);
          }
        \else
          \ifempty{#4}{
            \path[hgsem, \hgsrccol] (\n) -- (\hgname.center);
          }{
            \path[hgsem, \hgsrccol, #4] (\n) -- (\hgname.center);
          }
        \fi
      }
    \foreach \n in {#3}{
        \path (\n);
        \pgfgetlastxy \xntmp \yntmp;
        \pgfmathtruncatemacro{\xn}{\xntmp};
        \pgfmathtruncatemacro{\yn}{\yntmp};
        \pgfmathsetmacro\xtmpdiff{abs(\xn - \xcontrol)};
        \pgfmathsetmacro\ytmpdiff{abs(\yn - \ycontrol)};
        \ifdim \xtmpdiff pt > \hgtolerance
          \ifempty{#4}{
            \path[hgsem, ->, >=stealth', shorten >=1pt, \hgtgtcol] (\hgname.center) .. controls (tgtcentroid.center) and ($(tgtcentroid.center) + (tgtoffset)$) .. (\n);
          }{
            \path[hgsem, ->, >=stealth', shorten >=1pt, \hgtgtcol,#4] (\hgname.center) .. controls (tgtcentroid.center) and ($(tgtcentroid.center) + (tgtoffset)$) .. (\n);
          }
        \else
          \ifempty{#4}{
            \path[hgsem, ->, >=stealth', shorten >=1pt, \hgtgtcol] (\hgname.center) --  (\n);
          }{
            \path[hgsem, ->, >=stealth', shorten >=1pt, \hgtgtcol] (\hgname.center) --  (\n);
          }
        \fi
      }
    \fill[\hgfillcolor] (\hgname) circle [radius=\hgradius];
  }
  \endgroup
}

\newcommandx{\hgordeq}[1][1={\aH},usedefault=@]{\sqsubseteq_{#1}}
\newcommandx{\gintsem}[4][4=.5]{
  \tikz[hgsem,scale=#4,every node/.style={font=\scriptsize}]{
    \node (out) {$\aout[{#1}][{#2}][][{#3}]$};
    \node[below = 20pt of out] (in) {$\ain[{#1}][{#2}][][{#3}]$};
    \mkhg{out}{in};
  }
}

\newcommandx{\gsem}[2][1={\aG},2={},usedefault=@]{[\![ {#1} ]\!]_{#2}}
\newcommandx{\rbot}{\text{undef}}
\newcommandx{\rtrs}[1][1={\aH},usedefault=@]{{#1}^{\star}}
\newcommandx{\gord}[1][1={\aG},usedefault=@]{\leq_{#1}}
\newcommandx{\gordeq}[1][1={\aG},usedefault=@]{\leq_{#1}}
\mkfun{\cause}{cs}{}
\mkfun{\effect}{ef}{}
\newcommandx{\aW}{w}
\newcommandx{\rlang}{\mathcal{L}}
\newcommand{\gfun}[1]{\ensuremath{\mathsf{\colorFun #1}}}
\mkfun{\eact}{\gfun{act}}{}
\mkfun{\enode}{\gfun{cp}}{}

\mkuop{\rmax}{\gfun{max}}{\aH}
\mkuop{\rmin}{\gfun{min}}{\aH}
\mkuop{\rMAX}{\gfun{lst}}{\aH}
\mkuop{\rMIN}{\gfun{fst}}{\aH}

\newcommandx{\rseq}[2][1=\aG,2={\aG'},usedefault=@]{\gfun{seq}({#1},{#2})}
\newcommandx{\rpar}[2][1=\aG,2={\aG'},usedefault=@]{\gfun{par}({#1},{#2})}

\newcommandx{\gproj}[2][1=\aG,2=\ptp]{{#1}\downarrow_{#2}}
\newcommandx{\cinit}[1][1={\aQzero},usedefault=@]{{#1}}
\newcommandx{\cfinal}[1][1={q_e},usedefault=@]{{#1}}

\newcommandx{\geproj}[4][1=\aG,2=\ptp,3=\cinit,4=\cfinal,usedefault=@]{
{#1}\downarrow_{#2}^{{#3},{#4}}
}

\newcommand*{\StrikeThruDistance}{0.15cm}%
\tikzset{strike thru arrow/.style={
      decoration={markings, mark=at position 0.5 with {
              \draw [blue, thick,-]
              ++ (-\StrikeThruDistance,-\StrikeThruDistance)
              -- ( \StrikeThruDistance, \StrikeThruDistance);}
        },
      postaction={decorate},
    }}

\newcommandx{\ich}[1][1={\aG},usedefault=@]{{#1}^{\oplus}}
\newcommandx{\ichedges}[2][1={\aG},2={\gname},usedefault=@]{{#1}^{\oplus}({#2})}
\newcommandx{\parts}[1]{2^{#1}}
\newcommandx{\actch}{c}
\newcommandx{\soundactch}[2][1={\aG},2={\actch},usedefault=@]{{#1} \,\circledR\, {#2}}
\newcommandx{\rOnActch}[2][1={\aG},2={\actch},usedefault=@]{{#1} \setminus {#2}}
\newcommandx{\rOnActchClean}[2][1={\aG},2={\actch},usedefault=@]{{#1} \circledR {#2}}
\newcommandx{\rAllEvents}[1][1={\aG},usedefault=@]{\mathit{dom}(#1)}

\newcommand{\AV}{\mathcal{V}}
\newcommand{\aH}{H}

\newcommandx{\hgvertex}[2][1=\al,2=\gname,usedefault=@]{{#1}_{\textcolor{red}{[{#2}]}}}
\newcommand{\aE}{{\colorE E}}
\renewcommand{\ae}[1][e]{{\colorE{#1}}}

\newcommand{\al}[1][l]{{\colorE{#1}}}
\newcommandx{\hyedge}[1]{\{#1\}}

\newcommandx{\rdiv}[2][1=\gcho,2=\ptp,usedefault=@]{
  \gfun{div}_{#2}(#1)
}

\newcommandx{\rrdiv}[5][1={\aG},2={\aG'},3={\AV},4={,\AV'},5=\ptp,usedefault=@]{
  \gfun{div}^{#3#4}_{#5}(#1,#2)
}
\newcommandx{\pdiv}[3][1={\apom_1},2={\apom_2},3={\apom},usedefault=@]{
  \gfun{div}_{#3}(#1,#2)
}
\newcommandx{\pfork}[3][1={\apom_1},2={\apom_2},3={\apom},usedefault=@]{
  \gfun{fork}_{#3}(#1,#2)
}

\newcommandx{\mkint}[6][3=i,4=\p,5=\msg,6=\q,usedefault=@]{
  \node[bblock, #1] (#2) {$\gint[#3][#4][#5][#6]$};
}

\newcommand{\mkseq}[2]{\path[line] (#1) -- (#2);}

\newcommand{\mknseq}[1]{
  \StrCount{#1}{,}[\l] % from package xxstring
  \StrBefore{#1}{,}[\myhead]
  \StrBehind{#1}{,}[\mytail]
  \StrBefore{\mytail}{,}[\sndel]
  \ifnum \l > 1 {
        \mkseq{\myhead}{\sndel};
        \mknseq{\mytail}
      }
  \else{\ifnum \l > 0{
            \mkseq{\myhead}{\mytail};
          }
      \else{}
      \fi
    }
  \fi
}

\newcommandx{\mkgateblock}[6][6=yellow!10,usedefault=@]{
\path(#2);
\pgfgetlastxy{\xgate}{\ygate};
\pgfmathtruncatemacro{\xgateround}{\xgate};
\StrCount{#3,}{,}[\l] % from package xstring
\ifnum \l < 2 {\errmessage{#3 argument should be a comma-separated list of lenght >= 2}}
\else{
    \foreach \n in {#3}{
        \path (\n);
        \pgfgetlastxy{\xnode}{\ynode};
        \pgfmathtruncatemacro{\xnround}{\xnode};
        \pgfmathsetmacro\tmpdiff{abs(\xnround - \xgateround)}
        \ifdim \tmpdiff pt > 1 pt \path[line] (#2) -| (\n);
        \else
          \path[line] (#2) -- (\n);
        \fi
      }
  }
\fi
\StrCount{#4,}{,}[\l] % from package xstring
\ifnum \l < 2 {\errmessage{#4 argument should be a comma-separated list of lenght >= 2}}
\else{
    \foreach \n in {#4}{
        \path (\n);
        \pgfgetlastxy{\xnode}{\ynode};
        \pgfmathtruncatemacro{\xnround}{\xnode};
        \pgfmathsetmacro\tmpdiff{abs(\xnround - \xgateround)}
        \ifdim \tmpdiff pt > 1 pt \path[line] (\n) |- (#5);
        \else
          \path[line] (\n) -- (#5);
        \fi
      }
  }
\fi
\node[#1] at (#2) {};
\node[#1] at (#5) {};
{
\begin{pgfonlayer}{background}
  \path[fill=#6,rounded corners]
  (current bounding box.south west) rectangle
  (current bounding box.north east);
\end{pgfonlayer}
}
}

\newcommandx{\mkbranchblock}[5][5=@]{
  \mkgateblock{ogate}{#1}{#2}{#3}{#4}[#5]
}

\newcommandx{\mkforkblock}[5][5=@]{
  \mkgateblock{agate}{#1}{#2}{#3}{#4}[#5]
}

\newcommandx{\mkgraph}[3][1=.5cm, usedefault=@]{
  \node[source,above = #1 of {#2}] (src#2) {};
  \node[sink,below  = #1 of {#3}] (sink#3) {};
  \path[line] (src#2) -- (#2);
  \path[line] (#3) -- (sink#3);
}

\newcommandx{\mkloop}[5][1=.5, 2=1.5, 5=\aguard, usedefault=@]{
\node[lgate,above = #1 of {#3}] (entry#3) {};
\pgfgetlastxy \xentry \yentry;
\pgfmathtruncatemacro{\xentryrounded}{\xentry};
\node[below = #1 of {#4}, label = {above right:{$#5$}},yshift=-1em] (dummy) {};
\node[lgate,below  = #1 of {#4}] (exit#4) {};
\pgfgetlastxy \xexit \yexit;
\pgfmathtruncatemacro{\xexitrounded}{\xexit};
\path[line] (entry#3) -- (#3);
\path[line] (#4) -- (exit#4);
\pgfmathsetmacro\tmpdiff{abs(\xentryrounded - \xexitrounded)}
\path[line, color=teal] (exit#4) -| ($(exit#4)+(\tmpdiff,0)+(#2,0)$) |- (entry#3);
}

\newcommandx{\mkfork}[4][2=gatenode,3=i,4=.6,usedefault=@]{
  \mkgatebegin{#1}[{\gname[{#3}]}][agate][#4]{#2}
}

\newcommandx{\mkbranch}[4][2=gatenode,3=i,4=.6,usedefault=@]{
  \mkgatebegin{#1}[{\gname[{#3}]}][ogate][#4]{#2}
}

\newcommandx{\mkgatebegin}[5][2={},3=ogate,4=.5,usedefault=@]{
  \coordinate (gatecord) at (0,0);
  \coordinate (xmax) at (0,0);
  \coordinate (xmin) at (0,0);
  \pgfgetlastxy \xmin \xmax;
  \foreach \n [count=\i] in {#1}{
      \pgfgetlastxy \xc \yc;
      \path (\n);
      \pgfgetlastxy \xn \yn;
      \ifnum \i = 1
        \coordinate (xmin) at (\xn,0);
        \coordinate (xmax) at (\xn,0);
        \coordinate (max) at (0,\yn);
      \else
        \ifdim \xn < \xmin
          \coordinate (xmin) at (\xn,0);
        \fi
        \ifdim \xn > \xmax
          \coordinate (xmax) at (\xn,0);
        \fi
        \ifdim \yn < \yc
          \coordinate (max) at (0,\yc);
        \else
          \coordinate (max) at (0,\yn);
        \fi
      \fi
    }
  \coordinate (gatecord) at ($(xmin)!.5!(xmax) + (max) + (0,#4) + (max)$);
  \node[#3,label={below:$#2$}] (#5) at (gatecord) {};
  \pgfgetlastxy{\xgate}{\ygate};
  \pgfmathtruncatemacro{\xgateround}{\xgate};
  \StrCount{#1,}{,}[\l] % from package xxstring
  \ifnum \l < 2 {\errmessage{#1 argument should be a comma-separated list of lenght >= 2}}
  \else{
      \foreach \n in {#1}{
          \path (\n);
          \pgfgetlastxy{\xnode}{\ynode};
          \pgfmathtruncatemacro{\xnround}{\xnode};
          \pgfmathsetmacro\tmpdiff{abs(\xnround - \xgateround)}
          \ifdim \tmpdiff pt > 1 pt \path[line] (#5) -| (\n);
          \else
            \path[line] (#5) -- (\n);
          \fi
        }
    }
  \fi
}

\newcommandx{\mkgatebeginold}[5][2={},3=ogate,4=.5,usedefault=@]{
  \coordinate (gatecord) at (0,0);
  \foreach \n [count=\i] in {#1}{
      \pgfgetlastxy \xc \yc;
      \path (\n);
      \pgfgetlastxy \xn \yn;
      \coordinate (gatecord) at ($(gatecord) + (\xn,0)$);
      \coordinate (gatecord) at ($1/\i*(gatecord)$);
      \ifdim \yn < \yc
        \node (max) at (0,\yc) {};
      \else
        \node (max) at (0,\yn) {};
      \fi
    }
  \coordinate (gatecord) at ($(gatecord) + (0,#4) + (max)$);
  \node[#3,label={below:$#2$}] (#5) at (gatecord) {};
  \pgfgetlastxy{\xgate}{\ygate};
  \pgfmathtruncatemacro{\xgateround}{\xgate};
  \StrCount{#1,}{,}[\l] % from package xxstring
  \ifnum \l < 2 {\errmessage{#1 argument should be a comma-separated list of lenght >= 2}}
  \else{
      \foreach \n in {#1}{
          \path (\n);
          \pgfgetlastxy{\xnode}{\ynode};
          \pgfmathtruncatemacro{\xnround}{\xnode};
          \pgfmathsetmacro\tmpdiff{abs(\xnround - \xgateround)}
          \ifdim \tmpdiff pt > 1 pt \path[line] (#5) -| (\n);
          \else
            \path[line] (#5) -- (\n);
          \fi
        }
    }
  \fi
}

\newcommandx{\mkmerge}[4][2=gatenode,3=i,4=.5,usedefault=@]{
  \mkgateend{#1}[{\ifempty{#3}{}{\nmerge[#3]}}][ogate][#4]{#2}
}

\newcommandx{\mkjoin}[4][2=gatenode,3=i,4=.5,usedefault=@]{\mkgateend{#1}[{\ifempty{#3}{}{\nmerge[#3]}}][agate][#4]{#2}}

\newcommandx{\mkgateend}[5][2={},3=ogate,4=.5,usedefault=@]{
  \coordinate (gatecord) at (0,0);
  \coordinate (xmax) at (0,0);
  \coordinate (xmin) at (0,0);
  \pgfgetlastxy \xmin \xmax;
  \foreach \n [count=\i] in {#1}{
      \pgfgetlastxy \xc \yc;
      \path (\n);
      \pgfgetlastxy \xn \yn;
      \ifnum \i = 1
        \coordinate (xmin) at (\xn,0);
        \coordinate (xmax) at (\xn,0);
        \coordinate (ymin) at (0,\yn);
      \else
        \ifdim \xn < \xmin
          \coordinate (xmin) at (\xn,0);
        \fi
        \ifdim \xn > \xmax
          \coordinate (xmax) at (\xn,0);
        \fi
        \ifdim \yn > \yc
          \coordinate (ymin) at (0,\yc);
        \else
          \coordinate (ymin) at (0,\yn);
        \fi
      \fi
    }
  \coordinate (gatecord) at ($(xmin)!.5!(xmax) + (ymin)$);
  \node[#3,label={above:$#2$}] (#5) at ($(gatecord) - (0,{#4})$) {};
  \pgfgetlastxy{\xgate}{\ygate};
  \pgfmathtruncatemacro{\xgateround}{\xgate};
  \StrCount{#1,}{,}[\l] % from package xstring
  \ifnum \l < 2 {\errmessage{#1 argument should be a comma-separated list of lenght >= 2}}
  \else{
      \foreach \n in {#1}{
          \path (\n);
          \pgfgetlastxy{\xnode}{\ynode};
          \pgfmathtruncatemacro{\xnround}{\xnode};
          \pgfmathsetmacro\tmpdiff{abs(\xnround - \xgateround)}
          \ifdim \tmpdiff pt > 1 pt \path[line] (\n) |- (#5);
          \else
            \path[line] (\n) -- (#5);
          \fi
        }
    }
  \fi
}

\newcommandx{\mkgateendold}[5][2={},3=ogate,4=.5,usedefault=@]{
  \coordinate (gatecord) at (0,0);
  \coordinate (xmax) at (0,0);
  \coordinate (xmin) at (0,0);
  \pgfgetlastxy \xmin \xmax;
  \foreach \n [count=\i] in {#1}{
      \pgfgetlastxy \xc \yc;
      \path (\n);
      \pgfgetlastxy \xn \yn;
      \ifdim \xn < \xmin
        \coordinate (xmin) at (\xn,0);
      \fi
      \ifdim \xn > \xmax
        \coordinate (xmax) at (\xn,0);
      \fi
      \ifdim \yn > \yc
        \coordinate (ymin) at (0,\yc);
      \else
        \coordinate (ymin) at (0,\yn);
      \fi
      \coordinate (gatecord) at ($(xmin)!.5!(xmax) + (ymin)$);
    }
  \node[#3,label={above:$#2$}] (#5) at ($(gatecord) - (0,{#4})$) {};
  \pgfgetlastxy{\xgate}{\ygate};
  \pgfmathtruncatemacro{\xgateround}{\xgate};
  \StrCount{#1,}{,}[\l] % from package xstring
  \ifnum \l < 2 {\errmessage{#1 argument should be a comma-separated list of lenght >= 2}}
  \else{
      \foreach \n in {#1}{
          \path (\n);
          \pgfgetlastxy{\xnode}{\ynode};
          \pgfmathtruncatemacro{\xnround}{\xnode};
          \pgfmathsetmacro\tmpdiff{abs(\xnround - \xgateround)}
          \ifdim \tmpdiff pt > 1 pt \path[line] (\n) |- (#5);
          \else
            \path[line] (\n) -- (#5);
          \fi
        }
    }
  \fi
}

\newcommand{\gatedistancein}{3pt}
\newcommand{\gatedistanceinand}{2pt}

\usetikzlibrary{
  arrows,
  backgrounds,
  chains,
  calc,
  decorations.markings,decorations.pathreplacing,
  fadings,
  fit,
  patterns,
  petri,
  positioning,
  shadows,
  shapes,automata,shapes.callouts
}

\newcommand{\orgateG}{
  \begin{tikzpicture}[smallglobal,baseline=-.5ex, scale=0.75, every node/.style={transform shape}]
    \node [ogate] (o) {};
  \end{tikzpicture}
}

\newcommand{\pomset}[1]{
  \left[
	 \begin{tikzpicture}[pomset]
		#1
	 \end{tikzpicture}
  \right]
}

\tikzset{
  pomset/.style={
      node distance = .6cm and .6cm,
      scale = .7,
      transform shape,
      smooth
    }
}

\tikzset{
  src/.style={draw,circle,fill=white,
      minimum size=2mm,
      inner sep=0pt
    },
  sink/.style={draw,circle,double,fill=white,
      minimum size=1.5mm,
      inner sep=0pt
    },
  node/.style={draw,circle,fill=black,
      minimum size=2mm,
      inner sep=0pt
    },
  source/.style={draw,circle,fill=white,
      minimum size=3mm,
      inner sep=0pt
    },
  sink/.style={draw,circle,double,fill=white,
      minimum size=3mm,
      inner sep=0pt
    },
  block/.style = {rectangle, draw=gray, align=center, fill=orange!25, rounded corners=0.1cm,
      minimum size=5mm, inner sep=2pt},
  prenode/.style = {minimum size=9pt,inner sep=2pt, font=\Large},
  bblock/.style = {rectangle, draw=blue!50, opacity=.7, line width=.5pt, align=center, fill=white, rounded corners=0.1cm,
      minimum size=4mm, inner sep=1pt},
  prenode/.style = {minimum size=9pt,inner sep=2pt, font=\Large},
  agate/.style={draw, rectangle,
      minimum size=3mm,
      inner sep=0pt,
      fill=orange!25,
      label={[red]center:$\mid$}
    },
  ogate/.style = {
      diamond, draw, fill=orange!25,
      minimum size=4mm,
      inner sep=0pt,
      label={[red]center:$+$}
    },
  lgate/.style = {
      diamond, draw, fill=orange!25,
      minimum size=4mm,
      inner sep=0pt,
      label={[red]center:$\circlearrowleft$}
    },
  altogate/.style = {
      diamond, draw,
      minimum size=4mm,
      inner sep=0pt,
      postaction={path picture={% 
              \draw
              ([yshift=\gatedistancein]path picture bounding box.south) -- ([yshift=-\gatedistancein]path picture bounding box.north)
              ([xshift=-\gatedistancein]path picture bounding box.east) -- ([xshift=\gatedistancein]path picture bounding box.west)
              ;}}},
  altgate/.style={draw, rectangle,
      minimum size=3mm,
      inner sep=0pt,
      postaction={path picture={% 
              \draw
              ([yshift=\gatedistanceinand]path picture bounding box.south) --
              ([yshift=-\gatedistanceinand]path picture bounding box.north) ;}}},
  anygate/.style = {circle, draw, fill=white,
      minimum size=4mm,
      inner sep=0pt,
      postaction={path picture={% 
              \draw[black]
              ([xshift=-\gatedistancein,yshift=\gatedistancein]path picture bounding box.south east) --
              ([xshift=\gatedistancein,yshift=-\gatedistancein]path picture bounding box.north west)
              ([xshift=-\gatedistancein,yshift=-\gatedistancein]path picture bounding box.north east) --
              ([xshift=\gatedistancein,yshift=\gatedistancein]path picture bounding box.south west)
              ;}}
    },
  smallglobal/.style={
      node distance=1cm and 0.8cm, semithick, scale=0.8, every node/.style={transform shape}
    },
  elli/.style = {draw,densely dotted,-},
  line/.style = {draw,->, rounded corners=0.07cm,>=latex},
  nline/.style = {draw,semithick, ->},
  pline/.style = {draw,->,>=latex},
  node distance=1cm and 0.7cm,
  baseline=(current  bounding  box.center),
  local/.style={rectangle, draw, fill=\fillcolor, drop shadow,
      text centered, rounded corners, minimum height=5em
    },
  bigar/.style={
      draw,very thick, ->
    },
  process/.style={rectangle, draw=gray, fill=\fillcolor, drop shadow,
      text centered, minimum height=5em,text=gray
    },
  choreo/.style={rectangle, draw, fill=\fillcolor, drop shadow,
      text centered, rounded corners, minimum height=5em
    },
  mycfsm/.style={
      font=\footnotesize,
      initial where=above,
      ->,>=stealth,auto, node distance=1cm and 1cm,
      scale=1, every node/.style={transform shape},
      every state/.style=inner sep=2pt,
      baseline=(current  bounding  box.center),
      initial text={}
    },
  machinecloud/.style={
      cloud, cloud puffs=10, cloud ignores aspect, minimum height=.1cm, minimum width=2cm, draw
    },
  fitting node/.style={
      inner sep=0pt,
      fill=none,
      draw=none,
      reset transform,
      fit={(\pgf@pathminx,\pgf@pathminy) (\pgf@pathmaxx,\pgf@pathmaxy)}
    },
  mypetri/.style={
      font=\footnotesize,
      baseline=(current  bounding  box.center)
    },
  silentrans/.style = {rectangle, draw=black, align=center, fill=black,
      minimum height=1pt,
      minimum width=15pt,
      inner sep=1.5pt
    },
  reset transform/.code={\pgftransformreset},
  tmtape/.style={draw,minimum size=1.2cm}
}

\newcommand{\gunlessop}{\mbox{\colorOp\tiny\tt unless}}

\newcommandx{\gtry}[5][1=\gname,2={\aG_1 \gchoop \cdots \gchoop \aG_n},3=\gin,4=\gout,5={j},usedefault=@]{
\def\foo{\gtryop\ {#2} \ \gcatchop\ {#3} {\colorOp \Rightarrow} {#4} {\colorOp \bullet} {\gname[{#5}]}}
\gnode[{#1}][{\ifempty{#1} {\foo } {(\foo)}}]
}

\newcommandx{\gtrycatch}[4][1=\gname,2={\aG},3=\gin,4={\aG'},usedefault=@]{
\def\foo{\gtryop\ {#2} \ \gcatchop\ {#3} \gdoop\ {#4}}
\gnode[{#1}][{\ifempty{#1} {\foo} {(\foo)}}]
}

\newcommandx{\agG}[2][1={\aG},2=\aguard]{{#1} \ifempty{#2}{}{\ \gunlessop\ {#2}}}

\newcommandx{\grcho}[5][1=\gname,2={\agG},3={\agG[\aG'][\aguard']},4={\cdots},5=A,usedefault=@]{
\def\foo{{#2} {\ \ifempty{#4}{\gchoop}{\gchoop \ifempty{#4}{}{\ {#4}\  \gchoop}}\ } {#3}}
\ifempty{#1}{\ifempty{#5}{\foo}{\gselop\ \cpt[{#1}][{\ptp[#5]}]\big\{ \foo \big\}}}{\gselop\ \cpt[{#1}][{\ptp[#5]}]\big\{ \foo \big\}}
}

\newcommandx{\ggprefix}[3][1=\ptp,2={\aR},3={\aR'},usedefault=@]{f_{#1}} % it was \newcommandx{\common}{...}
\newcommand{\aconfigfn}{\chi}
\newcommand{\aconfig}{\ell}

\newcommand{\lstates}{\statemap}
\newcommandx{\sysconfig}[3][1=\lstates,2=\aconfigfn,3={},usedefault=@]{
  \conf{ {#1},{#2} \ifempty{#3}{}{, #3} }
}
\newcommand{\sysctxfn}[1][]{\gamma_{#1}}
\newcommandx{\sysctx}[2][1=\aQ,2={},usedefault=@]{({#1},\sysctxfn[{#2}])}

\newcommandx{\alog}[4][1=\msg,2=q,3=\gname,4=t,usedefault=@]{({#1},{#2},{#3},{#4})}
\newcommand{\aCM}{M}\newcommand{\aM}{\aCM}
\newcommand{\aQ}{Q}

\newcommandx{\aQzero}[1][1=,usedefault=@]{
  {\ifempty{#1}{q_0}{q_{0#1}}}
}
\newcommand{\badbranches}[1][]{\beta\ifempty{#1}{}{({#1})}}
\newcommandx{\guardedaction}[2][1=\al,2=\aguard,usedefault=@]{
  {#1} \ifempty{#2}{}{/} {#2}
}
\newcommandx{\atrM}[4][1=q,2=\al,3={\hat q,\hat \al, \aguard},4=q',usedefault=@]{
  {#1} \xrightarrow[{#3}]{\guardedaction[{#2}][]} {{#4}}
}
\newcommandx{\atrS}[5][
  1={\sysconfig[@][@][\badbranches]},
  2=\al,
  3=\aguard,
  4={\sysconfig[\lstates'][\aconfigfn'][\badbranches]},
  5=\sysctx,usedefault=@
]{
  {#1} \xRightarrow{\qquad} {{#4}}
}
\newcommandx{\arevtrS}[2][
  1={\sysconfig[@][@][\badbranches]},
  2={\sysconfig[\lstates'][\aconfigfn'][\badbranches']},
  usedefault=@
]{
  {#1} \rightsquigarrow {#2}
}
\newcommand{\aCS}{S}
\newcommand{\abuffer}{b}

\newcommandx{\enables}[2][1=\aconfigfn,2=\aguard,usedefault=@]{{#1} \vdash {#2}}
\newcommandx{\gprojfn}[5][1=\aG,2=\ptp,3=\cinit,4=\cfinal,5={},usedefault=@]{
  \mathbf{proj}_{#2}({#1},{#3},{#4}\ifempty{#5}{}{,{#5}})
}

\newcommandx{\rbp}[3][1=\aG,2=\aconfigfn,3=\achan,usedefault=@]{\mathtt{RBP}_{{#1},{#2}}\ifempty{#3}{}{({#3})}}

\newcommand{\apseudoCFSM}{\mathtt{M}}
\newcommandx{\pseudoseq}[2][1=\apseudoCFSM,2=\apseudoCFSM',usedefault=@]{{#1}  ; {#2}}
\newcommandx{\pseudoCFSM}[4][1=\aQ,2=\aQzero,3=\cfinal,4=\aTrs,usedefault=@]{(#1 \ ; #2 \ ; #3 \ ; #4)}
\newcommandx{\markt}[3][1=\hat{\al},2=\hat{q},3=\aguard,usedefault=@]{\%\big({#1} , {#2}, {#3}\big)}

\newcommandx{\borderfn}[2][1=\aconfig,2=\aloop,usedefault=@]{
\mathsf{border}_{{#2}}\ifempty{#1}{}{({#1})}
}

\newcommandx{\ggvisually}[6][1=5pt,2=15pt,3=5pt,4=5pt,5=1.0cm,6=\scriptsize,usedefault=@]{
  \def\dist{\hspace{#5}}
  \tikzset{
    mycallout/.style={
        fill=gray!10, opacity=.5, overlay, align=center,
        cloud callout, cloud puffs=15, aspect=1.9, cloud ignores aspect, cloud puff arc=100
      }
  }
  $\begin{array}{c@{\dist}c@{\dist}c@{\dist}c@{\dist}c}
      \begin{tikzpicture}[node distance=0.9cm and 0.4cm, every node/.style={scale=.7,transform shape}]
        \mkint{}{int}[]
        \mkgraph{int}{int};
        \node[mycallout, above = .3cm of srcint, xshift=1cm, callout absolute pointer={(srcint.east)}] {source node};
        \node[mycallout, below = .3cm of sinkint, xshift=-1cm, callout absolute pointer={(sinkint.west)}] {sink node};
      \end{tikzpicture}
       &
      \begin{tikzpicture}[node distance=.9cm and 0.4cm, every node/.style={scale=.7,transform shape}]
        \node[bblock] at (0,0) (g) {$\aG$};
        \node[node, below=of g] (s1) {};
        \node[bblock, below=of s1] (gp) {$\aG'$};
        \path[line,dotted] (g) -- (s1);
        \path[line,dotted] (s1) -- (gp);
      \end{tikzpicture}
       &
      \begin{tikzpicture}[node distance=.4cm and 0.4cm, every node/.style={scale=.7,transform shape}]
        \node[bblock] at (-.7,0) (g) {$\aG$};
        \node[bblock] at (.7,0)  (gp) {$\aG'$};
        \node[node, above=of g] (f) {};
        \node[node, below=of g] (j) {};
        \node[node, above=of gp] (fp) {};
        \node[node, below=of gp] (jp) {};
        \path[line,dotted] (f) -- (g);
        \path[line,dotted] (g) -- (j);
        \path[line,dotted] (fp) -- (gp);
        \path[line,dotted] (gp) -- (jp);
        \mkfork{f,fp}[fork][][#1];
        \mkjoin{j,jp}[join][][#2];
        \mkgraph{fork}{join};
        \node[mycallout, above = .3cm of fork, xshift=1cm, callout absolute pointer={(fork.east)}] {fork gate};
        \node[mycallout, above = -.9cm of join, xshift=-1cm, callout absolute pointer={(join.west)}] {join gate};
      \end{tikzpicture}
       &
      \begin{tikzpicture}[node distance=.4cm and 0.4cm, every node/.style={scale=.7,transform shape}]
        \node[bblock] at (-.7,0) (g) {$\aG$};
        \node[bblock] at (.7,0)  (gp) {$\aG'$};
        \node[node, above=of g] (f) {};
        \node[node, below=of g] (j) {};
        \node[node, above=of gp] (fp) {};
        \node[node, below=of gp] (jp) {};
        \path[line,dotted] (f) -- (g);
        \path[line,dotted] (g) -- (j);
        \path[line,dotted] (fp) -- (gp);
        \path[line,dotted] (gp) -- (jp);
        \mkbranch{f,fp}[fork][][#3];
        \mkmerge{j,jp}[join][][#4];
        \mkgraph{fork}{join};
        \node[mycallout, above = .3cm of fork, xshift=1cm, callout absolute pointer={(fork.east)}] {branch gate};
        \node[mycallout, above = -.9cm of join, xshift=-1cm, callout absolute pointer={(join.west)}] {merge gate};
      \end{tikzpicture}
       &
      \begin{tikzpicture}[node distance=0.4cm and 0.4cm, every node/.style={scale=.7,transform shape}]
        \node[bblock] (g) {$\aG$};
        \node[node, above=.5cm of g] (f) {};
        \node[node, below=.5cm of g] (j) {};
        \path[line,dotted] (f) -- (g);
        \path[line,dotted] (g) -- (j);
        \mkloop[.5][@]{f}{j}[];
        \mkgraph[.4cm]{entryf}{exitj};
        \node[mycallout, above = .2cm of entryf, xshift=1.3cm, callout absolute pointer={(entryf.east)}] {loop entry};
        \node[mycallout, above = -.7cm of exitj, xshift=-1.3cm, callout absolute pointer={(exitj.west)}] {loop exit};
      \end{tikzpicture}
      \\
      \text{#6 interaction}
       &
      \text{#6 sequential}
       &
      \text{#6 parallel}
       &
      \text{#6 branching}
       &
      \text{#6 iteration}
    \end{array}$
}

\date{}
\title{An Abstract Framework for Choreographic Testing\thanks{\tnxbehapi, \tnxitmatters[] and the \tnxtrustfull[]}}
\author{
  Alex Coto \institute{GSSI, Italy} \email{alex.coto@gssi.it}
  \and
  Roberto Guanciale \institute{KTH, Sweden}\email{robertog@kth.se}
  \and
  Emilio Tuosto \institute{GSSI, Italy and Univ. of Leicester, UK}\email{emilio.tuosto@gssi.it}
}
\def\authorrunning{A. Coto, R. Guanciale, E. Tuosto}
\def\titlerunning{An Abstract Framework for Choreographic Testing}

\graphicspath{{./figures/}{./diagrams/}}

\begin{document}
\maketitle

\begin{abstract}
  We initiate the development of a model-driven testing framework for
  message-passing systems.
  The notion of \emph{test} for communicating systems cannot simply be
  borrowed from existing proposals.
  Therefore, we formalize a notion of suitable distributed tests for a
  given choreography and devise an algorithm that generates tests as projections of
  global views.
  Our algorithm abstracts away from the actual projection operation,
  for which we only set basic requirements.
  The algorithm can be instantiated by reusing existing projection
  operations (designed to generate local implementations of global
  models) as they satisfy our requirements.
  Finally, we show the correctness of the approach and validate our methodology
  via an illustrative example.
\end{abstract}

\section{Introduction}\label{sec:intro}
We propose model-driven testing to complement the
\emph{correctness-by-construction} principle of choreographies.
We introduce a testing approach based on choreographies which we deem
% In fact, we consider choreographies particularly
suited to develop model-driven testing that may help to tame the
problems of correctness of distributed applications.

\paragraph{Context}
In the quest for correct-by-construction solutions, formal
choreographic models have proven themselves to be valuable approaches.
These models are gaining momentum, for instance, in the context of
business processes and message-passing applications.
The fundamental idea of choreographic models (originally proposed by
WS-CDL~\cite{w3c:cho}) is that specifications of systems consist of
\emph{global} and \emph{local views}.
The global view describes the behaviour of a system in terms of the
interactions among (the \emph{role}) of components.
The diagram below is an example of a global view of a protocol; we
will use this as a running example throughout the paper.
\\[.5pc]
\begin{tabular}{cc}
  \begin{minipage}[c]{.2\linewidth}
    \includegraphics[scale=.3]{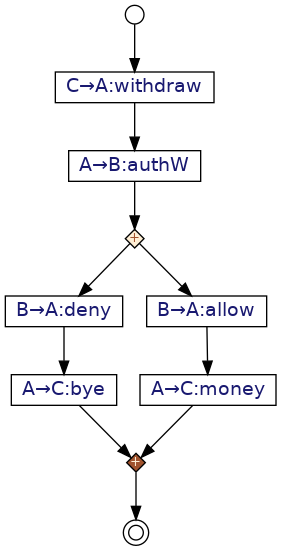}
  \end{minipage}
  &
  \begin{minipage}{.76\linewidth}
    This protocol is a simplified view of the main interactions that a
	 client \p[c] willing to withdraw some cash has to perform,
	 together with an ATM \p, and a bank \q.
	 The protocol starts with the \emph{interaction}
	 $\gint[][c][withdraw][a]$ with which participant \p[c] instructs
	 the ATM \p\ about the intention to withdraw some cash.
	 In the next interaction $\gint[][a][authW][b]$, \p\ asks the bank
	 \q\ to authorise the withdrawal.
	 Observe that payloads are abstracted away; for instance, the
	 message $\msg[withdraw]$ is intended to be a data type carrying
	 e.g., the amount of requested cash.
	 A distributed choice starts at the \emph{branching} point
	 $\orgateG$, where the bank \q\ decides whether to deny or grant
	 the withdrawal.
	 Note that the choice is non-deterministic since, besides from
	 data, this model abstracts away from local computations.
	 Depending on the local decision of \q, the next interaction is
	 either $\gint[][b][deny][a]$ or $\gint[][b][allow][a]$.
	 In each case the client is notified of the decision with
	 interactions $\gint[][a][bye][c]$ (in the first case) or
	 $\gint[][a][money][c]$ if the operation is granted by the bank.
  \end{minipage}
\end{tabular}
A main source of problems in distributed protocols is reaching
consensus among participants in distributed choices.
Indeed, participants have partial knowledge about the global
state of the protocol.
And, for the protocol to run \quo{smoothly}, the
partial knowledge of each participant should be consistent with respect to
the global state of the protocol.
For distributed choices this boils down to require \emph{awareness} of
each participant about the branch to follow.
For instance, in the example above, the bank is aware of the choice
since it decides what to do next and the other participants become
aware of the choice from the messages they exchange.

The correctness-by-construction principle of choreographic models is
usually realised through the identification of \emph{well-formedness}
conditions on global views. These are sufficient conditions guaranteeing
that the protocol can be executed distributively, without breaking the
consistency between the global state and the local knowledge of
participants.
In particular, formal choreographic approaches (such
as~\cite{hyc08,cdyp16,gt16,gt17,ivan,Bravetti2009} to mention a
few) study notions of well-formedness to guarantee the safety of
communications (usually, deadlock-freedom, no message losses, etc.).
The local view of a protocol indeed provides a specular specification
of the behaviour of (the role of) each component \quo{in isolation}.
In this way, the local view yields a set of computational units
enacting the communications specified in the global view.
For instance, the local view of the bank \q\ above consists of an
artefact waiting for a message $\msg[authW]$ from \p\ to which it
replies by sending either of the messages $\msg[deny]$ or
$\msg[allow]$.
Note that the client and the bank are \quo{oblivious} of each other, in the
sense that they interact only with the ATM.

The typical scheme to realise the correctness-by-construction
principle consists of the steps below:
\begin{steps}
  \item \label{enu:gg} provide an artefact defining the global view of
  the system;
  \item \label{enu:wf} revise the global view until well-formedness is
  achieved;
  \item \label{enu:prj} project global views into local views;
  \item \label{enu:|-} verify that code implementing the local view of a
  component complies with its projection.
\end{steps}
(It is also possible to avoid step~\ref{enu:|-} and project global
views directly on code.)
Steps~\ref{enu:gg} and~\ref{enu:wf} are mainly human activities,
although some algorithmic support\footnote{Some authors have
  considered the problem of supporting designers in the identification
  of problems in non well-formed
  choreographies~\cite{blt11,blt12,lmz13}.} is offered by the
verification of well-formedness conditions.
The remaining steps can instead be supported by algorithms.
In fact, (an approximation of) compliance is usually decidable and
projections can often be straightforwardly computed by \quo{splitting}
interactions into complementary send and receive actions.

\paragraph{Problem}
Although paramount for the development of message-passing
applications, the correctness-by-construction principle advocated by
formal choreographies is not enough.
At first sight this utterance may look controversial.
In fact, we do not contend that correctness-by-construction is not
worth pursuing (or not achievable: many models including those
mentioned above do realise the correctness-by-construction
principle).
But, even in a correctly implemented choreographic solution problems
may arise.
We list three major causes of possible disruption.
\begin{description}
  \item[Local computation] As said, formal choreographies focus on the
        interactions among components while abstracting away from local
        computations.
        Therefore, errors may still be introduced when developing code; for
        instance, a component expected to receive an integer and return a
        string, after inputting the integer may diverge on a local
        computation before delivering the expected string and cause a
        malfunction in the communication protocol.
  \item[Evolution] Software is often subject to continuous changes for
        instance to increase efficiency or to accommodate evolving
        requirements.
        For example, to reduce the communication overhead, a component may
        be modified so that two outputs are merged into one so to spare an
        interaction.
        Besides introducing bugs in the new code for local
		  computations, these changes may alter the original design
		  breaking the compliance required in step~\ref{enu:|-} of the
		  scheme above.
  \item[Openness] Increasingly, applications are built by composing
        computational elements developed independently and available
        off-the-shelf, over which the developer might have no control.
        This is for instance the main approach to develop service-oriented
        architectures.
        New releases or modifications of third-party components
		  (libraries, run-time support, etc.) may introduce malfunctions
		  in applications using it.
        For example, a new release of a service invoked by an application
        may enrich the spectrum of possible messages delivered to some
        components not designed to handle such new messages.
\end{description}

\paragraph{Contribution} We take a first step to equip known
choreography-based approaches with testing.
More precisely, we start addressing step~\ref{enu:prj} above.
Our main technical contribution is an algorithm to automatically
derive (abstract) test cases out of a well-formed choreography
(cf. \cref{sec:gen-algo}).
We develop our results in the setting of \emph{global
  choreographies}~\cite{gt16,gt18} and \emph{communicating finite
  state machines}~\cite{bz83}.
The former is the model we adopt to represent global views and the
latter is a well-known model for specifying communication protocols
that will serve to represent local views.

Our key contributions are:
\begin{itemize}
\item An abstract framework of well-formedness that captures the
  essential elements of formal choreographic models.
  This abstract framework makes our algorithm
  parametric with respect to the notion of well-formedness.
\item We lay down the definitions that transfer various notions of
  (standard) software testing to communication protocols.
  Formally this is done by adapting a few concepts from traditional
  software testing such as the notions of test~(\cref{def:testcase}),
  oracle~(\cref{def:scheme}), and test compliance~(\cref{def:compliance}).
  Again, the abstract framework paves the way for several alternative
  developments.
  We decided to explore one of them first; we discuss alternatives in
  \cref{sec:discussion}.
\item As we will see, not all test cases are \quo{meaningful},
  therefore we identify when tests
  are \emph{suitable} for a choreography   
  (\cref{def:testgood}).
\item We apply our framework to a non-trivial example
  (cf. \cref{sec:application}).
\end{itemize}

%%% Local Variables:
%%% mode: latex
%%% TeX-master: "main"
%%% End:

\section{Background}\label{sec:bkg}
We survey the main definitions and constructs needed in the rest of the paper.
We focus on \emph{global choreographies} (g-choreographies for short)
for the global view~\cite{gt17}, and borrow from~\cite{bz83} \emph{communicating
    finite-state machines} (CFSMs) for the local views.
G-choreographies were chosen because they offer an intuitive visual
description together with a precise semantics~\cite{gt18,gt19}.
We adopt CFSMs because they have many similarities with programming
languages based on message-passing, such as Erlang.

\subsection{Global Choreographies}\label{sec:gg}
The global view of a choreography can be suitably specified as
a \emph{global choreography}~\cite{gt18,gt16,dy12}.
This model is appealing as it has a syntactic and diagrammatic
presentation, and has been given a formal semantics in terms of pomsets, which
enable for automatic processing.

Fix a set $\ptpset$ of \emph{participants} and a set $\msgset$ of
\emph{message (types)} such that $\ptpset \cap \msgset = \emptyset$;
let $\p, \q, \ldots$ range over $\ptpset$ and $\msg, \msg[n], \ldots$
range over $\msgset$.
A global choreography (or g-choreography) is a term derivable from the following grammar:
\[\begin{array}{lcl@{\hspace{2cm}}r}
		\aG & \bnfdef & \gempty & \text{empty}
		\\  & \bnfmid & \gint & \text{interaction}
		\\[.1em]  & \bnfmid & \aG \gparop \aG & \text{fork}
		\\[.1em]  & \bnfmid & \aG \gchoop \aG & \text{choice}
		\\[.1em]  & \bnfmid &  \aG \gseqop \aG & \text{sequential}
		\\[.1em]  & \bnfmid &  \grepkw\ { \aG } & \text{iteration}
		% \\[.1em]  & \bnfmid & \gblk
	\end{array}
\]

The empty choreography $\gempty$ yields no interactions; trailing
occurrences of $\gempty$ may be omitted.
An interaction $\gint[]$ represents the exchange of a message of type
$\msg$ between $\p$ and $\q$, provided that $\p \neq \q$.
We remark that data are abstracted away: in $\gint[]$, the message
$\msg$ is not a value and should rather be thought of as (the name of)
a data type\footnote{We leave implicit the grammar of data types; in
	the examples we will assume that $\msg$ ranges over basic types such
	as $\msg[int]$, $\msg[bool]$, $\msg[string]$, etc.}.
G-choreographies can be composed  sequentially or in parallel
($\gseq[]$ and $\gpar[]$).
A (non-deterministic) choice
$\aG_1 \gchoop \aG_2$ specifies
the possibility to continue according to either $\aG_1$ or $\aG_2$.
The body $\aG$ in an iteration $\grepkw\ { \aG }$ is repeated
until a participant in $\aG$ (non-deterministically) chooses 
to exit the loop.
Although for simplicity we do not consider iterative g-choreographies
in our examples, the techniques we introduce further on can work on
arbitrary finite unfoldings of the loops, as is commonplace in
software testing or in verification techniques such as bounded
model-checking.

\begin{example}\label{ex:atmgg}
	The g-choreography for the example introduced in \cref{sec:intro} is
	\[
		\aG_{\text{ATM}} =
		\gseq[][{\gint[][c][withdraw][a]}][{
		  \gseq[][{\gint[][a][authW][b]}][{
			   \gblk[{\gseq[][{\gint[][b][deny][a]}][{\gint[][a][bye][c]}]}
			         \gchoop
			         {\gseq[][{\gint[][b][allow][a]}][{\gint[][a][money][c]}]}
			   ]
		  }]
		}]
	\]
	where we assume that sequential composition takes precedence
        over choice.
\end{example}

The semantics of a g-choreography as defined in~\cite{gt16,gt18} is a
family of pomsets (partially ordered multisets); each pomset in the
family is the partial order of events occurring on a particular
\quo{branch} of the g-choreography.
Events are therefore labelled by \emph{(communication) actions} $\al$
occurring in the g-choreography.
The output of a message $\msg \in \msgset$ from
participant $\p \in \ptpset$ to participant $\q \in \ptpset$ is
denoted by $\aout$, while the corresponding input is denoted by $\ain$.
More formally,
\begin{align*}
	\lact & = \{\aout, \ain \mid \p[A], \p[B] \in \ptpset \text{ and }  \msg \in \msgset\}
\end{align*}
is the set of (communication) \emph{actions} and $\al$ ranges over
$\lact$.
The subject of an action is defined as $\subject[\aout] = \p$ and
$\subject[\ain] = \q$.

It is not necessary to restate here the whole constructions for the
semantics which is given by induction on the structure of the
g-choreography; we simply give an informal account.
The semantics $\gsem[\gempty]$ is the set $\{\emptypom\}$ containing
the empty pomset $\emptypom$, while for interactions we have
\[
	\gsem[{\gint[]}] =
	\pomset{
		\node (out) at (0,0) {$\aout[@][@][]$};
		\node[right = of out] (in) {$\ain[@][@][]$};
		\path[draw,->] (out) -- (in);
	}
\]
namely, the semantics of an interaction is a pomset where the output
event precedes the input event.
The semantics of the other operations is basically obtained
by composing the semantics of sub g-choreographies.
More precisely,
\begin{itemize}
	\item for a choice we essentially have
	      $\gsem[{\gcho[]}] = \gsem \cup \gsem[\aG']$;
	\item the semantics of the parallel composition $\gpar[]$ is
	      essentially built by taking the disjoint union of each pomset in
	      $\gsem$ with each one in $\gsem[\aG']$;
	\item the semantics of the sequential composition $\gsem[{\gseq[]}]$
	      is the disjoint union of each pomset in $\gsem$ with each one in
	      $\gsem[\aG']$ and, for every participant $\ptp$, making every output
	      of $\ptp$ in $\gsem$ precede all events of $\ptp$ in $\gsem[\aG']$.
\end{itemize}
\begin{example}\label{ex:atmsem}
	Consider $\aG_{\text{ATM}}$ of \cref{ex:atmgg}.
	We have
	\[
		\gsem[{\aG_{\text{ATM}}}] = \left\{
		\begin{array}[c]{c}
			\pomset{
				\node (0) {$\aout[c][a][][withdraw]$};
				\node[right = of 0] (1) {$\ain[c][a][][withdraw]$};
				\node[right = of 1] (2) {$\aout[a][b][][authW]$};
				\node[right = of 2] (3) {$\ain[a][b][][authW]$};
				\node[right = of 3] (4) {$\aout[b][a][][deny]$};
				\node[right = of 4] (5) {$\ain[b][a][][deny]$};
				\node[right = of 5] (6) {$\aout[a][c][][bye]$};
				\node[right = of 6] (7) {$\ain[a][c][][bye]$};
				\foreach \i/\j in {0/1,1/2,2/3,3/4,4/5,5/6,6/7}{
						\path[->,draw] (\i) -- (\j);
					}
			},
			\\
			\pomset{
				\node (0) {$\aout[c][a][][withdraw]$};
				\node[right = of 0] (1) {$\ain[c][a][][withdraw]$};
				\node[right = of 1] (2) {$\aout[a][b][][authW]$};
				\node[right = of 2] (3) {$\ain[a][b][][authW]$};
				\node[right = of 3] (4) {$\aout[b][a][][allow]$};
				\node[right = of 4] (5) {$\ain[b][a][][allow]$};
				\node[right = of 5] (6) {$\aout[a][c][][money]$};
				\node[right = of 6] (7) {$\ain[a][c][][money]$};
				\foreach \i/\j in {0/1,1/2,2/3,3/4,4/5,5/6,6/7}{
						\path[->,draw] (\i) -- (\j);
					}
			}
		\end{array}
		\right\}
	\]
	For the sake of illustration, the singleton
	\[\left\{
		\pomset{
			\node (0) {$\aout[c][a][][withdraw]$};
			\node[right = of 0] (1) {$\ain[c][a][][withdraw]$};
			\node[right = of 1] (2) {$\aout[a][b][][authW]$};
			\node[right = of 2] (3) {$\ain[a][b][][authW]$};
			\node[above right = of 3] (4) {$\aout[b][a][][deny]$};
			\node[right = of 4] (5) {$\ain[b][a][][deny]$};
			\node[right = of 5] (6) {$\aout[a][c][][bye]$};
			\node[right = of 6] (7) {$\ain[a][c][][bye]$};
			\node[below right = of 3] (8) {$\aout[b][a][][allow]$};
			\node[right = of 8] (9) {$\ain[b][a][][allow]$};
			\node[right = of 9] (10) {$\aout[a][c][][money]$};
			\node[right = of 10] (11) {$\ain[a][c][][money]$};
			\foreach \i/\j in {0/1,1/2,2/3,3/4,4/5,5/6,6/7,3/8,8/9,9/10,10/11}{
					\path[->,draw] (\i) -- (\j);
				}
		}
		\right\}
	\]
	is the semantics of the g-choreography obtained by replacing choice
	with parallel composition in $\aG_{\text{ATM}}$.
\end{example}

The language of a g-choreography $\aG$, written $\rlang[\aG]$, is the
closure under prefix of the set of all \emph{linearizations} of
$\gsem[\aG]$ where a linearisation of a pomset is a permutation of its
events that preserves the order of the pomset.
\begin{example}\label{ex:atmL}
  The language of the last pomset in \cref{ex:atmsem}
  is the set of prefixes of words obtained by concatenating
$\aout[c][a][][withdraw]\ \ain[c][a][][withdraw] \
		\aout[a][b][][authW]\ \ain[a][b][][authW]$ with both $\aout[b][a][][deny]\
		\ain[b][a][][deny]\ $ $
		\aout[a][c][][bye]\
		\ain[a][c][][bye]$ and $
		\aout[b][a][][allow]\
		\ain[b][a][][allow]\ $ $ 
		\aout[a][c][][money]\
		\ain[a][c][][money]$.
\end{example}

%%% Local Variables:
%%% mode: latex
%%% TeX-master: "main"
%%% End:

\subsection{Communicating Systems}\label{sec:cs}
As in~\cite{lty15,dy12}, we adopt \emph{communicating finite state
  machines} (CFSMs) as local artefacts.
We borrow the definition of CFSMs in~\cite{bz83} adapting it to our
context.
A CFSM $\aCM = \aCFSM$ is a finite transition system where
\begin{itemize}
  \item $\aQ$ is a finite set of {\em states} with \emph{initial} state
        $\aQzero \in \aQ$, and
  \item $\aTrs\ \subseteq \ \aQ \times \lact \times \aQ$; we write
        $q \trans{\al} {q'}$ for $(q,\al,q') \in \aTrs$.
\end{itemize}
Machine $\aCM$ is \emph{local} to a participant $\p \in \ptpset$ (or
\emph{$\p$-local}) if $\subject[\al] = \p$ for each transition
$q \trans{\al} {q'}$ of $\aM$.
A \emph{(communicating) system} is a map
$\aCS = (\aCM_{\ptp})_{\ptp \in \ptpset}$ where
$\aCM_{\ptp} = \aCFSM[\aQ_{\ptp}][{\aQzero[\ptp]}][\aTrs_{\ptp}]$ is a
$\p$-local CFSM for each $\ptp \in \ptpset$.
The set of \emph{channels} (fixed for all communicating systems) is
$\chset = \{\achan \sst \p \neq \q \in \ptpset\}$; for all
$\achan \in \chset$, it is assumed that there is an unbound finite
multiset $\abuffer_{\p \q} = \multiset{\msg_1, \ldots, \msg_n}$
containing the messages that $\aCM_{\p}$ sends to $\aCM_{\q}$ and from
which $\aCM_{\q}$ consumes the messages sent by $\aCM_{\p}$.
We use $\_ \multicup \_$ for multiset union and $\_ \multidiff \_$ for
multiset difference.

The semantics of communicating systems is defined in terms of
\emph{transition systems}, which keep track of the state of each
machine and the content of each buffer.
Let $\aCS = (\aCM_{\ptp})_{\ptp \in \ptpset}$ be a communicating
system.
A \emph{configuration} of $\aCS$ is a pair
$\aConf = \csconf q \abuffer$ where
$\vec q = (q_{\p})_{\p \in \ptpset}$ with $q_{\p} \in \aQ_{\p}$ and
$\vec \abuffer = (\abuffer_{\achan})_{\achan \in \chset}$ mapping each
channel to a multiset of messages; $q_{\p}$ keeps track of the
\emph{local state} of machine $\aM_{\p}$ in $\aConf$ and buffer
$\abuffer_{\achan}$ keeps track of the messages sent from $\p$ to
$\q$.
The \emph{initial} configuration $\aConf_0$ is the one where, for all
$\p \in \ptpset$, $q_{\p}$ is the initial state of the corresponding
CFSM and all buffers are empty.

A configuration $\aConf'= \csconf{q'}{\abuffer'} $ is \emph{reachable}
from another configuration $\aConf = \csconf{q}{\abuffer}$ by
\emph{firing an $\al$-transition}, written
$\aConf \cstr{\al} \aConf'$, if there is a message $\msg \in \msgset$
such that either (1) or (2) below holds:
\begin{center}
  \begin{tabular}{l@{\hspace{.5cm}}r}
    \begin{minipage}{.45\linewidth}\small
      1.
      $\al = \aout$ , $\vec q(\p) \trans[\p]{\al} {\vec{q'}(\p)}$,
      and
      \begin{itemize}
        \item[a.] $\vec{q'} = \upd{\vec q}{\p}{\vec{q'}(\p)}$ and
        \item[b.] $\vec{\abuffer'} = \upd{\vec \abuffer} \achan {\vec \abuffer(\achan) \multicup \multiset \msg}$
      \end{itemize}
    \end{minipage}
     &
    \begin{minipage}{.45\linewidth}\small
      2.
      $\al = \ain$, $\vec q(\q) \trans[\q]{\al} \vec{q'}(\q)$,
      $\vec \abuffer (\achan)(\msg) > 0$,
      and
      \begin{itemize}
        \item[a.]  $\vec{q'} = \upd{\vec q}{\p}{\vec{q'}(\p)}$ and
        \item[b.] $\vec{\abuffer'} = \upd{\vec \abuffer} \achan
                {\vec \abuffer(\achan) \multidiff \multiset \msg}$
      \end{itemize}
    \end{minipage}
  \end{tabular}
\end{center}
(where $\upd f x y$ is the usual update operation that redefines
function $f$ on an element $x$ of its domain with $y$).
Condition (1) puts $\msg$ on channel $\ptp\ptp[B]$, while (2) gets
$\msg$ from channel $\ptp\ptp[B]$.
In both cases, any machine or buffer not involved in the transition
is left unchanged in the new configuration $\aConf'$.

Note that this construction differs from the original definition
in~\cite{bz83}, (where unbounded FIFO queues were used) in order to make
the communication model similar to the one of Erlang.

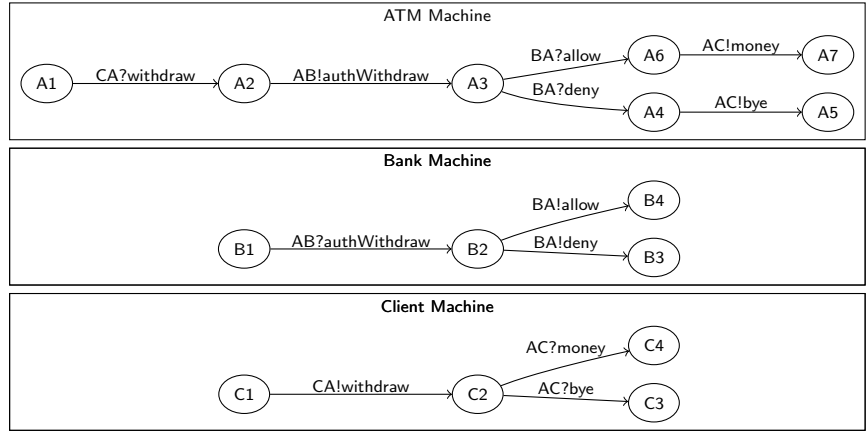
\begin{figure}
  \centering
  \begin{tikzpicture}[scale=0.4, every node/.style={scale=0.8}]
  \begin{dot2tex}[codeonly]
  digraph G {
    // pencolor=transparent

    rankdir=LR;
    compound=true;
    newrank=true;

    subgraph cluster_TC {
      label="Client Machine";
      C1 -> C2 [label="CA!withdraw"]
      C2 -> C3 [label="AC?bye"]
      C2 -> C4 [label="AC?money"]
      
      { // Extend box size
        Co[style=invis];
        C0[style=invis];
        edge[style=invis]
        C0 -> C1
        C4 -> Co
      }
    }

    subgraph cluster_TB {
      label="Bank Machine";

      B1 -> B2 [label="AB?authW"]

      B2 -> B3 [label="BA!deny"]
      B2 -> B4 [label="BA!allow"]

      { // Extend box size
        B0[style=invis]
        Bo[style=invis]
        edge[style=invis]
        B0 -> B1
        B4 -> Bo
      }
    }

    subgraph cluster_TA {
      label="ATM Machine";

      A1 -> A2 [label="CA?withdraw"]

      A2 -> A3 [label="AB!authW"]
      A3 -> A4 [label="BA?deny"]
      A4 -> A5 [label="AC!bye"]

      A3 -> A6 [label="BA?allow"]
      A6 -> A7 [label="AC!money"]
    }

    {   // Alignment
        rank=same; A2; C1; B1;
        // { edge [style=invis]; A2->C1->B1; }
    }
  }
  \end{dot2tex}
\end{tikzpicture}
  \caption{CFSMs for the protocol in \cref{sec:intro}}
  \label{fig:atm-minimal-CFSMs}
\end{figure}

\begin{example}
  A local view of the protocol in \cref{sec:intro} is given by the
  CFSMs in \cref{fig:atm-minimal-CFSMs}.
  Notice how the events reflected in the global view have been split
  into their \emph{send} and \emph{receive} counterparts.

  The starting state of each CFSM is the leftmost one.
  The CFSM of the client initiates the protocol by sending a
  withdraw message to the ATM, which reacts by sending a message to
  the bank to check whether the client can actually perform this
  withdrawal.
  CFSMs \p\ and \q\ will gradually proceed as they take messages out
  from the queues existing between all pairs of participants.

  State $\mathsf{B2}$ of \q\ is the \emph{internal choice} state that
  corresponds to the branching point of the g-choreography.
  Namely, in $\mathsf{B2}$, the bank locally chooses how to proceed.
  As soon as \q\ sends either an $\msg[allow]$ or a $\msg[deny]$
  message, the ATM either delivers the money or finishes the conversation 
  with a $\msg[bye]$ message.
\end{example}

A configuration $\aConf = \csconf q \abuffer$ is \emph{stable} if all
buffers are empty (note that stability does not impose any requirement
on a machine's enabled transitions): $\aConf$ is \emph{stable} for
$\chset' \subseteq \chset$ if all buffers in $\chset'$ are empty in
$\aConf$, and it is a \emph{deadlock} if $\aConf \not{\cstr{}}$ and
either there is a participant $\ptp \in \ptpset$ such that
$\vec q(\p) \trans[\p] \ain$ or $\aConf$ is not stable.
This definition is adapted from~\cite{cf05} and is meant
to capture communication misbehaviour.
Observe that, according to this definition, a configuration $\aConf$
where all machines are in a state with no outgoing transitions and all
buffers are empty is not a deadlock configuration even though
$\aConf \not{\cstr{}}$.

Let $\mrun[\aCS,\aConf]$ be the set of \emph{runs} of a communicating
system $\aCS$ starting from a configuration $\aConf$ of $\aCS$, that
is the set of sequences
$\arun = \{(\hat\aConf_i, \al_i, \hat\aConf_{i+1})\}_{0 \leq i \leq
  n}$ with $n \in \nat \cup \{\infty\}$ such that
$\hat\aConf_0 = \aConf$, and
$\hat\aConf_i \cstr{\al_i} \hat\aConf_{i+1}$ for every
$0 \leq i \leq n$; we say that run $\arun$ is \emph{maximal} if
$n = \infty$ or $\hat\aConf_{n} \not{\cstr{\ }}$ and denote with
$\mrun[\aCS]$ the runs of $\aCS$ starting from its initial state.
The \emph{language of a communicating system} $\aCS$ is the set
\[
  \rlang[\aCS] = \bigcup_{\arun \in \mrun[\aCS]} \{\text{trace of } \arun\}
\]
where the \emph{trace} of a run
$\{(\hat\aConf_i, \al_i, \hat\aConf_{i+1})\}_{0 \leq i \leq n} \in
  \mrun[\aCS]$ is the sequence $\al_0 \dots \al_{n-1}$.
Notice that $\rlang[\aCS] \subseteq \lact^\omega \cup \lact^\star$ (where $\lact^\omega$ is the set of infinite words over $\lact$) and
it is prefix-closed.

%%% Local Variables:
%%% mode: latex
%%% TeX-master: "main"
%%% End:

\section{Generating Tests}\label{sec:generation}
The goal of model-driven testing is to find mismatches between a
specification and an implementation.
We focus on \emph{component-testing}, which in our setting corresponds
to test a single participant of a g-choreography.  We dub
\emph{component under test} (CUT) an implementation which should be
tested.

\subsection{Baseline concepts}
Top-down approaches of choreographies define projection functions that
generate local models from global models.
In order to parameterise our framework with respect to these notions,
we introduce \emph{abstract projections} on g-choreographies.
\begin{definition}[Abstract projection]\label{def:abstract}
  A map $\projpom{\_}{\_}$ is an \emph{abstract projection} if it
  takes a g-choreography $\aG$ and a participant $\p \in \ptpset$ and
  returns an \p-local CFSM.
  Given a g-choreography $\aG$, the \emph{system induced by
	 $\projpom{\_}{\_}$} is defined as
  $\projpom{\aG}{} = (\projpom{\aG}{\p})_{\p \in \ptpset}$.
\end{definition}
There are several ways to define projection operations that are
instances of \cref{def:abstract}.
For example, in~\cite{gt16,gt18} a g-choreography is projected on a
participant \p\ in two steps, which we briefly summarise since we will
illustrate our framework by adopting this operation in our examples.
By induction on the structure of the g-choreography, the first step
transforms each interaction in the transition of an automaton
according to the role of \p\ in that interaction.
More precisely, the interaction becomes an output or an input
transition depending on whether \p\ is the sender or the receiver;
otherwise the iteration corresponds to a silent transition.
In the second step, the CFSM obtained as above is determinised.
\begin{example}
  The CFSMs shown in \cref{fig:atm-minimal-CFSMs} are obtained by
  means of the projection operation in~\cite{gt16,gt18} applied to the
  g-choreography $\aG_{\text{ATM}}$ in \cref{ex:atmgg} where some
  equivalent states (e.g., $\mathsf{A5}$ and $\mathsf{A7}$) are
  replicated for readability.
\end{example}

Not every g-choreography can be \emph{faithfully} projected.
In fact, the asynchronous semantics of communicating systems may
introduce behaviour that does not correspond to the intended behaviour
of the g-choreography.
In concrete instances, sufficient conditions on g-choreographies are
given so that the semantics of projected communicating systems reflect
the semantics of the g-choreography.
These conditions are abstractly captured in the next definition.
\begin{definition}[Abstract well-formedness]\label{def:awf}
  A predicate on g-choreographies is an \emph{abstract well-formedness
	 condition} if $\wellformed[\aG]$ implies that there is a
  communicating system $\aCS$ with initial configuration $\aConf_0$
  such that
    \begin{itemize}
        \item $\rlang[\aCS] \subseteq \rlang[\aG]$ and
        \item no run in $\mrun(\aConf_0)$ contains a deadlock configuration;
    \end{itemize}
    in this case we say that $\aCS$ \emph{realises} $\aG$.
\end{definition}
Note that \cref{def:awf} admits trivial instances such as the
predicate which does not hold on any g-choreography.
The choreography in \cref{sec:intro} is considered well-formed in the
majority of existing work.
In this example there is only one participant that makes a choice
(i.e., the bank) and the rest of the participants are informed of
which decision was taken.
Intuitively, this avoids coordination problems and therefore the
choreography can be correctly realized by CFSMs, such as the ones in
\cref{fig:atm-minimal-CFSMs}.
In this case, the language of the choreography is the same as the
language of the projected communicating system, which is
deadlock-free.

Hereafter, we assume projections that \emph{respect} abstract
well-formedness.
\begin{definition}[Compatible projections]\label{def:compatible}
    An abstract projection $\projpom{\_}{\_}$ is \emph{compatible with
        $\wellformed$} when, for all g-choreographies $\aG$, if
    $\wellformed[\aG]$ then the system induced by $\projpom{\_}{\_}$
    realises $\aG$.
\end{definition}
An abstract projection mapping all participants to a machine without
any transitions is trivially compatible with any abstract well-formedness
condition.
Of course, we are interested in abstract projections for which
$\rlang[\projpom{\aG}{}] = \emptyset$ only if $\aG = \gempty$.

We can now formalise the main notions of our choreographic testing
framework.
A \emph{test case} for a CUT \p\ is a set of CFSMs with a
distinguished set of success states; the outcome of a test case is
determined by its interaction with \p.
\begin{definition}[Test case]\label{def:testcase}
  A \emph{test case for a CUT $\p \in \ptpset$} is a set
  $\atestcase = \{\conf{\aM_1, \Qsucc_1}, \ldots,
  \conf{\aM_n,\Qsucc_n}\}$ such that for every $1 \leq i,j \leq n$,
  $\aM_{i} = (\aQ_{i}, \aQzero_{i}, \aTrs_{i})$ is a CFSM with
  $\Qsucc_{i} \subseteq \aQ_{i}$ and
    \begin{itemize}
        \item if $q \cfsmtr[\al]_{i} q'$ then $\subject[\al] \neq \p$
			 \hfill (1)
        \item if $q \cfsmtr[{\aout[b][c]}]_{i} q'$ and
			 $q \cfsmtr[\al]_{i} q''$ then $\al = \aout[b][c]$
			 \hfill (2)
        \item if $q \cfsmtr[\al]_{i} q'$ and $q \cfsmtr[\al]_{i} q''$
			 then $q' = q''$
			 \hfill (3)
        \item if $q_1 \cfsmtr[\al]_i q_2$ and
			 $q_1' \cfsmtr[\al']_j q_2'$ and $\subject[\al] =
			 \subject[\al']$ then $i = j$ \hfill (4)
    \end{itemize}
    We dub $\Qsucc_{i}$ the \emph{success states} of $\aM_{i}$.
\end{definition}

  We briefly justify the conditions in \cref{def:testcase}.
  Condition (1) forces the CUT not to be the subject of any
  transition, since tests cannot force it directly to take specific actions.
  Conditions (2) and (3) together enforce that there is always a
  single possible output for the system to proceed, that the
  machines are deterministic and, in particular, that they cannot have
  internal choice or mixed\footnote{A mixed choice state is one with
	 both input and output outgoing transitions.} states.
  The rationale behing conditions (2) and (3) is to \quo{confine}
  non-determinism in the CUT and its concurrent execution with the
  test so that it is easier to analyse the outcome of tests.
  The last condition enforces transitions across machines to have
  different subjects: if this was not the case, generating code for
  each participant could be significantly more complex.
  Note that this does not force the CFSMs in a test case to be necessarily local; in
  fact, \cref{def:testcase} admits different subjects in the labels of
  different transitions.  

The following example shows the requirements of \cref{def:testcase} and a violation of those requirements.
\begin{example}\label{ex:validtest}
  Consider $\aM_{\ptp[a]}$, $\aM_{\ptp[b]}$ and $\aM_{\ptp[c]}$ in
  \cref{fig:atm-minimal-CFSMs} that respectively are the CFSMs of the
  ATM, the bank, and the client.
  Then
  $\atestcase_1 = \{ \conf{\aM_{\ptp[A]}, \left\{ \mathsf{A5},
		  \mathsf{A7} \right\}}, \conf{\aM_{\ptp[C]}, \left\{ \mathsf{C3},
		  \mathsf{C4} \right\}} \}$ is a test case for $\q$ (i.e.,
  bank).
  In fact, $\aM_{\ptp[A]}$ and $\aM_{\ptp[C]}$ are deterministic,
  internal choice-free and do not include any transitions where the
  subject is \q.
  Instead,
  $\atestcase_2 = \{ \conf{\aM_{\ptp[B]}, \left\{ \mathsf{B3},
		\mathsf{B4} \right\}}, \conf{\aM_{\ptp[C]}, \left\{ \mathsf{C3},
		\mathsf{C4} \right\}} \}$ is not a test case for $\p$ (i.e., the
  ATM) because $\aM_{\ptp[B]}$ has an internal choice in state
  $\mathsf{B2}$.
  \end{example}

\newcommand{\testcompCh}[3]{#1 \triangleright_{#3} #2}
\newcommand{\testcompAll}[2]{#1 \triangleright #2}
\begin{definition}[Test compliance]\label{def:compliance}
  Let $\chset' \subseteq \chset$ be a set of channels, $\hat \aM$ a
  CFSM, and $\atestcase$ a test case.
  Denote with $\testsys[\hat \aM]$ the communicating system
  consisting of $\hat \aM$ and the CFSMs in $\atestcase$.
  We say that $\hat \aM$ is \emph{$\atestcase$-compliant w.r.t $\chset'$}
  ($\testcompCh{\hat \aM}{\atestcase}{\chset'}$) if every finite
  maximal run of $\testsys[\hat \aM]$ contains a stable configuration $\aConf$
  for $\chset'$ such that for every
  $\conf{\aM, \Qsucc} \in \atestcase$ the local state of $\aM$ in
  $\aConf$ is in $\Qsucc$.
    % A successful configuration is stable, and every machine has to be in a successful state. 
\end{definition}
In the following, we dub the configuration $\aConf$ in
\cref{def:compliance} a \emph{successful configuration for $\atestcase$}
and we use $\testcompAll{\hat \aM}{\atestcase}$ for
$\testcompCh{\hat \aM}{\atestcase}{\chset}$.
Notice that the parametrization on $\chset'$ allows a CFSM to be
considered compliant even if some runs leave channels in
$\chset \setminus \chset'$ not empty.
The next series of examples illustrate the notion of test compliance
with four tests for CUTs in \cref{fig:atm-minimal-CFSMs}.
\begin{example}
  Let $\q$ be the CUT and $\atestcase_1$ be the test case in
  \cref{ex:validtest}.
  Then $\aM_{\q}$ is $\atestcase_1$-compliant.
  In fact, the system consisting of $\aM_{\q}$ and (the CFSMs in)
  $\atestcase_1$ is exactly the system implementing the choreography
  of the running example.
  However, $\aM_{\q}$ is not compliant with the test case
  $\{ \conf{\aM_{\p}, \left\{ \mathsf{A3} \right\}},
  \conf{\aM_{\ptp[C]}, \left\{ \mathsf{C3}, \mathsf{C4} \right\}} \}$.
  In fact, \p[c] can reach $\mathsf{C3}$ or $\mathsf{C4}$ only after
  that \p\ has left state $\mathsf{A3}$.
  Similarly, $\aM_{\q}$ is not compliant with the test case
  $\{\conf{\aM_{\p}, \left\{ \mathsf{A7} \right\}},
	 \conf{\aM_{\ptp[C]}, \left\{ \mathsf{C3} \right\}} \}$, since the
  success states of \p\ and \p[C] represent conflicting branches.
\end{example}

\begin{example}
  Suppose that the CUT is the CFSM $\aM'_{\q}$ obtained by removing
  the transition $\aout[\q][\p][][allow]$ from $\aM_{\q}$.
  Then $\testcompAll{\aM'_{\q}}{\atestcase_{1}}$ however, $\aM'_{\q}$
  is not compliant with
  $\{\conf{\aM_{\p}, \left\{ \mathsf{A7} \right\}},
  \conf{\aM_{\ptp[C]}, \left\{ \mathsf{C4} \right\}} \}$.
  This is due to the fact that the test and the CUT select different
  branches.
  Similarly, $\aM'_{\q}$ is not compliant with
  $\{ \conf{\aM'_{\p}, \left\{ \mathsf{A7} \right\}},
	 \conf{\aM_{\ptp[C]}, \left\{ \mathsf{C3}, \mathsf{C4} \right\}}
  \}$, where $\aM'_{\p}$ is obtained by removing the transition
  $\ain[\q][\p][][deny]$ from $\aM_{\p}$.
\end{example}

\begin{example}
  Finally, let $\p$ be the CUT and $\aM'_{\q}$ be the CFSM obtained by
  removing the transition $\aout[\q][\p][][allow]$ from $\aM_{\q}$.
  Then $\aM_{\p}$ is compliant with
  $\{ \conf{\aM'_{\q}, \left\{ \mathsf{B3} \right\}},
	 \conf{\aM_{\ptp[C]}, \left\{ \mathsf{C3}, \mathsf{C4} \right\}}
  \}$.
\end{example}

\newcommand{\good}{\text{suitable}}
\newcommand{\adherent}{\text{adherent}}
We finally define when a test case is meaningful for a choreography, by requiring that the correct implementation (i.e., the projection) of the choreography is compliant with the test.

\begin{definition}[Test suitability]\label{def:testgood}
  Test $\atestcase$ is $(\aG,\p)-\good$ if
  $\projpom\aG\p \triangleright \atestcase$.
\end{definition}

\subsection{Test generation algorithm}
\label{sec:gen-algo}

\mkfun{\gtst}{split}{}
\mkfun{\gtstset}{tests}{}
\mkfun{\alternatives}{\textit{nds}}{}
\newcommand{\testgen}[2]{\textit{tests}(#1,#2)}

To generate tests we follow a straightforward strategy: we start from
the projections of the participants that are not the CUT and we remove their internal choices.
The intuition is that for well formed g-choreographies, the
projections are \quo{compatible} with any implementation that restricts
internal choices with respect to the projection of the CUT.
We use the following auxiliary function to identify non-deterministic
states.
These are the states that the algorithm uses to split the transitions
to obtain deterministic tests.
Given a CFSM $\aM = (\aQ, \aQzero, \aTrs)$, let
\[
  \alternatives[\aM] = \left\{q \sst \exists q \cfsmtr[\al_1] q_1 \neq q
	 \cfsmtr[\al_2] q_2 \qst \al_1 = \al_2 
	 \vee \{\al_1,\al_2\} \cap \in \lact^! \neq \emptyset) \right\}
\]
be the set of non-deterministic states of $\aM$, that is the states
with at least two different transitions that either have the same
label or one of which is an output transition.
For convenience, we let $\aM(q)$ denote the set of outgoing
transitions of $q$ in $\aM$ and $\aM - t$ (resp. $\aM + t$) be the
operation that removes from (resp. adds to) $\aM$ transition $t$
(these operations extend element-wise to sets of transitions).
The following function produces sets of machines that are
internal choice free:
\begin{align*}
  \gtst[\aM] =
  &
  \begin{cases}
    \{\aM\} & \text{if } \alternatives[\aM] = \emptyset
    \\
    \bigcup_{q \in \alternatives[\aM]} \gtst[\aM, q] & \text{otherwise}
  \end{cases}
  & 	  \quad  \qquad
  \\
  \gtst[\aM, q] =
  &
	 \begin{cases}
		\displaystyle{\bigcup_{q \cfsmtr[\aout] q'}}
		\gtst[{\aM - \aM(q) + q \cfsmtr[\aout] q'}] & \text{if } \aM(q) \text{ has output transitions}
		\\[2em]
		\displaystyle{\bigcup_{\substack{q \cfsmtr[\ain] q'
				\\ \neq \\
				q \cfsmtr[\ain] q''
			 }
		  }
		}
	 \gtst[{\aM - q \cfsmtr[\ain] q'}] & \text{otherwise}
  \end{cases}
\end{align*}

Once these simpler CFSMs are obtained, success states have to be set
for each of them.
This is analogous to problem commonly known in software testing as the
\emph{oracle problem}: deciding when a test is successful.
This decision is application-dependent and its solutions usually
requires human intervention~\cite{bhmsy15}.
In our setting, this corresponds to single out configurations of
communicating systems according to a sub-tree of a choreography
as defined below.
Intuitively, we would like success states from the CFSMs to correspond to the execution of specific syntactic subtrees of the choreography.

We now introduce an additional definition that helps us determine
the success states for our tests.
In the following, given a g-choreography $\aG$, let $\subgraphs[\aG]$ be the set of
sub-trees of the abstract syntax tree producing $\aG$ once we fix a
suitable precedence among the operators.
Our algorithm relies on abstract syntax trees of g-choreographies,
but it does not depend on the precedence relation chosen.

\begin{definition}[Oracle scheme]\label{def:scheme}
    Let $\aG$ be a g-choreography, $\projpom{\_}{\_}$ an abstract
    projection compatible with a given well-formedness condition
    $\wellformed[]$.
    An \emph{oracle scheme of $\aG$} for $\projpom{\_}{\_}$ is a
    function $\termstates_{\aG,\ \projpom{}{}}$ mapping a pair
    $(\p,\atree) \in \ptpset \times \subgraphs[\aG]$ on a set of states
    of the CFSM $\projpom \aG \p$ such that if $\wellformed[\aG]$ 
    and $\aCS$ is the communicating system induced by $\projpom{\_}{\_}$, then
    for every $\atree \in \subgraphs[\aG]$ and maximal run $\pi \in
        \mrun[\aCS]$ there exists a stable configuration $\aConf$ in $\pi$
    such that, for each $\ptp \in \ptpset$, for the local state $q_{\p}$
    of $\ptp$ in $\aConf$ we have that $q_{\ptp} \in
        \termstates_{\aG,\ \projpom{}{}}(\ptp,\atree)$.
\end{definition}
The main purpose of the oracle scheme $\termstates_\aG$ is to map
a participant and a subtree $(\p,\atree) \in \ptpset \times \subgraphs[\aG]$ 
to a set of states of $\projpom{\aG}{\ptp}$ that correspond to the states 
the system can be in after the execution of the sub-tree $\atree$ of $\aG$.

\begin{example}
  Below is a fragment of a possible oracle scheme for the g-choreography from
  \cref{ex:atmgg} and the CFSMs shown in
  \cref{fig:atm-minimal-CFSMs}.
  \[
	 \termstates_{\aG,\ \projpom{}{}} (\p, \aG) = \{\mathsf{A5}, \mathsf{A7}\}
	 \qqand
	 \termstates_{\aG,\ \projpom{}{}} (\q, \aG) = \{\mathsf{B3}, \mathsf{B4}\}
	 \qqand
	 \termstates_{\aG,\ \projpom{}{}} (\p[c], \aG) = \{\mathsf{C3}, \mathsf{C4}\}
  \]
  \[
	 \termstates_{\aG,\ \projpom{}{}} (\p, \gint[][B][allow][A]) = \{\mathsf{A6}, \mathsf{A4}\}
	 \text{ and }
	 \termstates_{\aG,\ \projpom{}{}} (\q, \gint[][B][allow][A]) = \{\mathsf{B3}, \mathsf{B4}\}
	 \text{ and }
	 \termstates_{\aG,\ \projpom{}{}} (\p[c], \gint[][B][allow][A]) = \{\mathsf{C3}, \mathsf{C4}\}
  \]
  Notice that for the whole g-choreography $\aG$, the oracle scheme
  $\termstates_{\aG,\ \projpom{}{}}$ yields the last states of the
  CFSMs, and for the sub-tree $\gint[][B][allow][A]$ it returns the
  first state that allows the participant to acknowledge either the
  execution of the interaction or the selection of an alternative
  branch.
\end{example}

Test cases are then built by combining machines obtained by the $\gtst$
function and by identifying the success states via the oracle function, i.e. states that correspond to
the execution of the interactions of the subtrees of the g-choreography:
\begin{align}
  \testgen{\aG}{\p} =
  \left\{
  (\conf{\aM_{\q}, \termstates_{\aG,\ \projpom{}{}}(\q,\atree)})_{\q \neq \p \in \ptpset}
  \sst \forall \q \neq \p \in \ptpset \qst \aM_{\q} \in \gtst[\projpom \aG \q] \land \atree \in \subgraphs(G)
  \right\}
  \label{eq:testgen}
\end{align}

More intuitively, for every participant we select a single machine 
from the ones generated by $\gtst$ and combine them (exhaustively) into test cases.
Each test case corresponds to a unique path of execution
(i.e. selection of internal choices) of the original g-choreography.

\begin{theorem}\label{thm:correctness}
  If $\wellformed[\aG]$ then every test case in $\testgen \aG \p$ is
  $(\aG,\p)-\good$.
  \label{thm:gen-tests-are-good}
\end{theorem}

\newcommand{\merged}{\otimes}
\newcommand{\runs}[1]{\Pi\left(#1\right)}
\newcommand{\aRun}{\pi}
\newcommand{\aCUT}{\ptp[A]}
\newcommand{\aProj}{\downharpoonright}
\newcommand{\aTestSystem}{\aG\aProj_{\aCUT} \merged \atestcase}
\newcommand{\aState}{\mathsf{q}}

%%% Local Variables:
%%% mode: latex
%%% TeX-master: "main"
%%% End:

\section{Choregraphy-based Testing}\label{sec:application}
\begin{figure}[hbp]
    \centering
    \includegraphics[scale=.35]{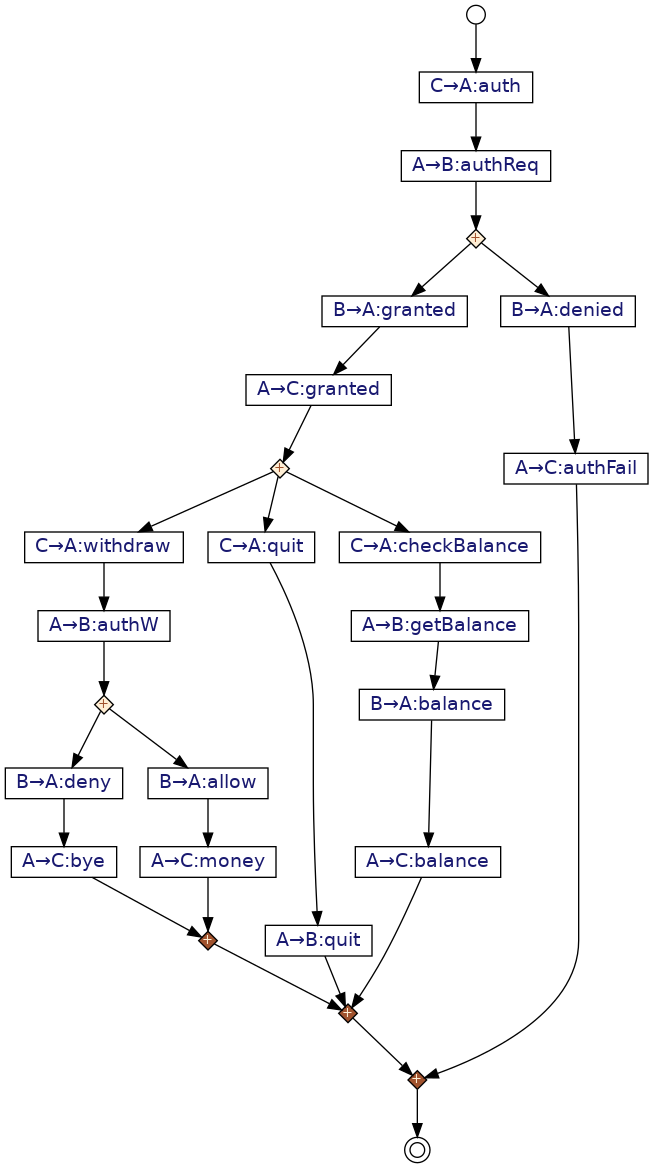}
    \caption{The complete choreography for the ATM scenario}
    \label{fig:atm-global}
\end{figure}

We now delve into a larger example in order to demonstrate the test
generation procedure in a more complex scenario.
\cref{fig:atm-global} shows a choreography involving the participants
\p\, \q, and \p[C], i.e., respectively the ATM, the bank, and a client
as in the running example used so far.
Observe that the running example is a sub-choreography of the
g-choreography in \cref{fig:atm-global}.
The bigger scenario can be straightforwardly represented as a
g-choreography as done in \cref{ex:atmgg} for the choreography in
\cref{sec:intro}.

The client starts a session of the protocol by authenticating with the
ATM machine ($\msg[auth]$).
The ATM then delegates the authentication to the bank, which can
either reject or accept the request by replying with either a
$\msg[denied]$ or a $\msg[granted]$ message.
In both cases the ATM forwards the authentication result to the
client.
The choreography terminates if the authentication fails.
If authentication is successful then the ATM offers three options to
the client: (M) withdraw money ($\msg[money]$), (Q) terminate the
session ($\msg[quit]$), or (B) check the account balance
$\msg[checkBalance]$.

In case (B), the ATM requests to the bank the balance and forwards the
result to the client via a $\msg[balance]$ message.
In case (Q), the ATM simply notifies the bank of the termination of
the session.
Case (M) is the choreography of \cref{sec:intro} whereby the
withdrawal request is forwarded to the bank which decides if to allow
or deny the request.

We demonstrate the test case generation for the ATM (i.e., participant
\p\ in \cref{fig:atm-global} is our CUT).
We first project the g-choreography corresponding to the choreography
in \cref{fig:atm-global}, using again the projection operation
in~\cite{gt18}.
We obtain the three CFSMs of~\cref{fig:atm-projections}.
The oracle scheme is visually represented by decorating states
only for two sub-trees of the choreography. More precisely:
\begin{itemize}
\item double-circles denote the states marked by the oracle scheme for
  the whole choreography, and
\item gray-circles correspond to the states marked by the oracle
  scheme for the interaction  $\gint[][B][allow][A]$.
\end{itemize}
It is straightforward to check that the system
consisting of these three CFSMs effectively generates the same
language as the one generated by the g-choreography.

\begin{figure}
    \centering
    \begin{tikzpicture}[scale=0.3, every node/.style={scale=0.6}]
    \begin{dot2tex}[codeonly]
        digraph G {
        rankdir=LR;
        node[label=""];

        subgraph cluster_TC {
        label="Client Machine";

        C1 -> C2 [label="CA!auth"]
        C2 -> C3 [label="AC?authFail"]
        C2 -> C4 [label="AC?granted"]

        C4 -> C5 [label="CA!withdraw"]
        C4 -> C6 [label="CA!quit"]
        C4 -> C7 [label="CA!checkBalance"]

        C5 -> C8 [label="AC?bye"]
        C5 -> C9 [label="AC?money"]

        C7 -> C10 [label="AC?balance"]

            {
                // Final states
                node[shape=doublecircle];
                C10; C6; C8; C9; C3;
            }

            {   // Extend box size
                Co[style=invis]
                edge[style=invis]
                C10 -> Co
            }

            {
                // Internal choice
                node[fillcolor = red, style=filled]
                C4;
            }

            {
                // Allow branch
                node[fillcolor = lightgray, style=filled]
                C3; C8; C9; C7;
            }

        }

        subgraph cluster_TB {
        label="Bank Machine";

        B1 -> B2 [label="AB?authReq"]

        B2 -> B3 [label="BA!denied"]
        B2 -> B4 [label="BA!granted"]

        B4 -> B5 [label="AB?authW"]
        B4 -> B8 [label="AB?getBalance"]
        B4 -> B10 [label="AB?quit"]

        B5 -> B6 [label="BA!deny"]
        B5 -> B7 [label="BA!allow"]

        B8 -> B9 [label="BA!balance"]

            {
                // Final states
                node[shape=doublecircle];
                B9; B7; B6; B3; B10;
            }

            {
                // Internal choice
                node[fillcolor = red, style=filled]
                B2; B5;
            }

            {   // Extend box size
                Bo[style=invis]
                edge[style=invis]
                B9 -> Bo
            }

            {
                // Allow branch
                node[fillcolor = lightgray, style=filled]
                B7; B6; B8; B3
            }

        }

        subgraph cluster_TA {
        label="ATM Machine";

        A1 -> A2 [label="CA?auth"]
        A2 -> A3 [label="AB!authReq"]
        A3 -> A4 [label="BA?denied"]
        A4 -> A5 [label="AC!authFail"]

        A3 -> A6 [label="BA?granted"]
        A6 -> A7 [label="AC!granted"]

        A7 -> A8 [label="CA?withdraw"]
        A7 -> A14 [label="CA?quit"]
        A7 -> A16 [label="CA?checkBalance"]

        A8 -> A9 [label="AB!authW"]
        A9 -> A10 [label="BA?deny"]
        A10 -> A11 [label="AC!bye"]

        A9 -> A12 [label="BA?allow"]
        A12 -> A13 [label="AC!money"]

        A14 -> A15 [label="AB!quit"]

        A16 -> A17 [label="AB!getBalance"]
        A17 -> A18 [label="BA?balance"]
        A18 -> A19 [label="AC!balance"]

            {rank=same; A6; A1;}

            {
                // Final states
                node[shape=doublecircle];
                A5; A15; A19; A11; A13;
            }

            {
                // Allow branch
                node[fillcolor = lightgray, style=filled]
                A4; A16; A14; A12; A10;
            }

        }

        }
    \end{dot2tex}
\end{tikzpicture}
    \caption{Projections of the choreography of \cref{fig:atm-global}} 
    \label{fig:atm-projections}
\end{figure}

\newcommand{\accentcolor}{ red }

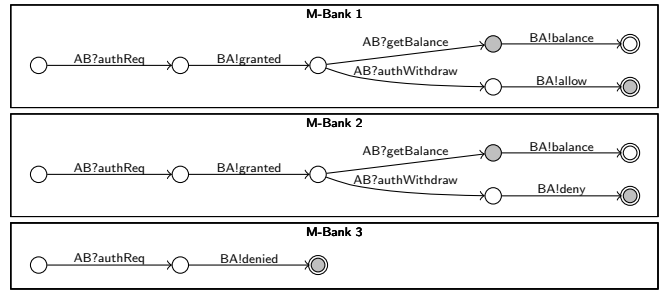
\begin{figure}
    \centering
    \begin{tikzpicture}[scale=0.3, every node/.style={scale=0.6}]
    \begin{dot2tex}[codeonly]
        digraph G {
        rankdir=LR;
        node[label=""];

        subgraph cluster_TB_1 {
        label="M-Bank 3"
        Ba1 -> Ba2 [label="AB?authReq"]

        // Internal choice 1
        Ba2 -> Ba3 [label="BA!denied"]

        // Ba4 -> Ba5 [label="AB?authW"]
        // Ba4 -> Ba8 [label="AB?getBalance"]

        // // Internal choice 2
        // Ba5 -> Ba6 [label="BA!deny"]

        // Ba8 -> Ba9 [label="BA!balance"]

        {
            // Final state
            Ba3[shape=doublecircle];
        }

        {
            // Marked states for B->A:allow
            Ba3[fillcolor = lightgray, style=filled];
        }

        {   // Box enlargement
            Ba3 -> Ba4 -> Ba5 [style=invis]
            Ba4[style=invis]
            Ba5[style=invis]
        }

        }

        // After trimming, bank 2 = bank 1
        // subgraph cluster_TB_2 {
        // label="Bank 2"
        // Bb1 -> Bb2 [label="AB?authReq"]

        // // Internal choice 1
        // Bb2 -> Bb3 [label="BA!denied"]

        // // Bb4 -> Bb5 [label="AB?authW"]
        // // Bb4 -> Bb8 [label="AB?getBalance"]

        // // // Internal choice 2
        // // Bb5 -> Bb7 [label="BA!allow"]

        // // Bb8 -> Bb9 [label="BA!balance"]

        // }

        subgraph cluster_TB_3 {
        label="M-Bank 2"
        Bc1 -> Bc2 [label="AB?authReq"]

        // Internal choice 1
        Bc2 -> Bc4 [label="BA!granted"]

        Bc4 -> Bc5 [label="AB?authW"]
        Bc4 -> Bc8 [label="AB?getBalance"]

        // Internal choice 2
        Bc5 -> Bc6 [label="BA!deny"]

        Bc8 -> Bc9 [label="BA!balance"]

        {
            // Final states
            node[shape=doublecircle];
            Bc6; Bc9;
        }

        {
            // Marked states for B->A:allow
            Bc8[fillcolor = lightgray, style=filled];
            Bc6[fillcolor = lightgray, style=filled];
        }

        }

        subgraph cluster_TB_4 {
        label="M-Bank 1"
        Bd1 -> Bd2 [label="AB?authReq"]

        // Internal choice 1
        Bd2 -> Bd4 [label="BA!granted"]

        Bd4 -> Bd5 [label="AB?authW"]
        Bd4 -> Bd8 [label="AB?getBalance"]

        // Internal choice 2
        Bd5 -> Bd7 [label="BA!allow"]

        Bd8 -> Bd9 [label="BA!balance"]

        {
            // Final states
            node[shape=doublecircle];
            Bd7; Bd9;
        }

        {
            // Marked states for B->A:allow
            Bd8[fillcolor = lightgray, style=filled];
            Bd7[fillcolor = lightgray, style=filled];
        }

        }

        }
    \end{dot2tex}
\end{tikzpicture}
    \caption{CFSMs resulting from splitting the projected CFSMs for the bank}
    \label{fig:atm-splitted-bank}
\end{figure}

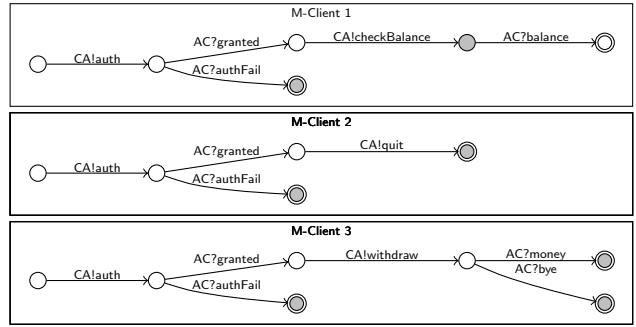
\begin{figure}
    \centering
    \begin{tikzpicture}[scale=0.3, every node/.style={scale=0.6}]
    \begin{dot2tex}[codeonly]
        digraph G {
        rankdir=LR;
        node[label=""];

        subgraph cluster_TC_1 {
        label="M-Client 3"
        Ca1 -> Ca2 [label="CA!auth"]
        Ca2 -> Ca3 [label="AC?authFail"]
        Ca2 -> Ca4 [label="AC?granted"]

        Ca4 -> Ca5 [label="CA!withdraw"]

        Ca5 -> Ca8 [label="AC?bye"]
        Ca5 -> Ca9 [label="AC?money"]
        // Removed with trimming
        // Ca7 -> Ca10 [label="AC?balance"]

        {
            // Final states
            node[shape=doublecircle];
            Ca3; Ca9; Ca8;
        }

        {
            // Marked states for B->A:allow
            Ca3[fillcolor = lightgray, style=filled];
            Ca8[fillcolor = lightgray, style=filled];
            Ca9[fillcolor = lightgray, style=filled];
        }

        }

        subgraph cluster_TC_2 {
        label="M-Client 2"
        Cb1 -> Cb2 [label="CA!auth"]
        Cb2 -> Cb3 [label="AC?authFail"]
        Cb2 -> Cb4 [label="AC?granted"]

        Cb4 -> Cb6 [label="CA!quit"]
        // Removed with trimming
        // Cb5 -> Cb8 [label="AC?bye"]
        // Cb5 -> Cb9 [label="AC?money"]

        // Cb7 -> Cb10 [label="AC?balance"]

        {
            // Final states
            node[shape=doublecircle];
            Cb6; Cb3;
        }

        {
            // Marked states for B->A:allow
            Cb3[fillcolor = lightgray, style=filled];
            Cb6[fillcolor = lightgray, style=filled];
        }

        {   // Box enlargement
            Co[style=invis]
            Cb6 -> Co[style=invis]
        }

        }

        subgraph cluster_TC_3 {
        label="M-Client 1"
        Cc1 -> Cc2 [label="CA!auth"]
        Cc2 -> Cc3 [label="AC?authFail"]
        Cc2 -> Cc4 [label="AC?granted"]

        Cc4 -> Cc7 [label="CA!checkBalance"]
        // Removed with trimming
        // Cc5 -> Cc8 [label="AC?bye"]
        // Cc5 -> Cc9 [label="AC?money"]

        Cc7 -> Cc10 [label="AC?balance"]
        }

        {
            // Final states
            node[shape=doublecircle];
            Cc10; Cc3;
        }

        {
            // Marked states for B->A:allow
            Cc3[fillcolor = lightgray, style=filled];
            Cc7[fillcolor = lightgray, style=filled];
        }

        }
    \end{dot2tex}
\end{tikzpicture}
    \caption{CFSMs resulting from splitting the projected CFSMs for the client}
    \label{fig:atm-splitted-client}
\end{figure}

The two machines for $\ptp[B]$ and $\ptp[C]$ of
\cref{fig:atm-projections} cannot be directly used as a test for $\p$
since they have states with internal choices.
These states, obtained by applying $\alternatives$ to the
CFSMs of \q\ and \p[c], are the sets of red states shown in the figure.
At this point the algorithm applies $\gtst[\q]$ to compute a set of
four machines, say $\mathbb{M}_{\q}$. This is done by selecting in all
possible ways one of the output transitions from states of \q\
(according to the second clause in the definition of $\gtst[]$).
Likewise for $\p[c]$, the algorithm produces a set of three
machines, say $\mathbb{M}_{\p[c]}$.
The resulting sets of CFSMs are shown in
\cref{fig:atm-splitted-bank,fig:atm-splitted-client} where, for the
sake of conciseness, we remove unreachable states, also omitting
isomorphic CFSMs.
For a sub-tree of the choreography, 
we obtain a test case by combining a machine from $\mathbb{M}_{\q}$ and one from
$\mathbb{M}_{\p[c]}$ and defining their success states using the oracle scheme.
Function $\testgen{\aG}{\p}$ generates all the test cases by freely choosing the machines as above and 
exhaustively iterating over the sub-trees of the choreography.

This process results in nine tests for each sub-tree of the
g-choreography.
For the tree corresponding to the whole g-choreography, the success
states are those depicted as double-circles.
For the tree corresponding to the interaction $\gint[][B][allow][A]$,
the success states are those in gray.
Notice that some states are success states for both trees.
Moreover, all the resulting tests satisfy the requirements of
\cref{def:testcase}.

%%% Local Variables:
%%% mode: latex
%%% TeX-master: "main"
%%% End:

\section{Discussion \& Open Problems}\label{sec:discussion}
We started the exploration of mechanisms to support model-driven
testing of message-passing systems based on choreographies.
To this purpose, we decided to rely on the so called top-down approach
featured by an existing choreographic model.
The choreographic model adopted here is rather abstract, but it
is close to real programming paradigms such as those of Erlang.
We exploited the notion of projection of global views of
choreographies in order to devise an automatic test generation
mechanism.
The design of our algorithm required us to fix the basic notion of
test, test feasibility, and test success within the framework of
g-choreographies and communicating systems.
Although we tried to give a general framework that abstracts away from
actual projection operations, we took some design decisions for the
identification of our framework.

The notion of test case considered here (\cref{def:testcase}) requires
tests not to contain mixed-choice states (that is, states with both
output and input outgoing transitions).
In fact, without assumptions on the projection operation
mixed-choice states cannot be split easily as they are.
Consider the system consisting of following CFSMs:
\begin{align}\label{eq:mixed}
  \aM_{\p} =
  \begin{tikzpicture}[cfsm]
    \node[state] (0) {$q_0$};
    \node[draw=none,fill=none] (start) [left  = 0.3cm  of 0]{};
    \node[state] (1) [below of=0]   {$q_1$};
    \path  (start) edge node {} (0)
    (0) edge node[above] {$\aout$} (1)
    ;
  \end{tikzpicture}
  \quad\qquad
  \aM_{\q} =
  \begin{tikzpicture}[cfsm]
    \node[state] (0) {$q_0$};
    \node[draw=none,fill=none] (start) [left  = 0.3cm  of 0]{};
    \node[state] (1) [below left of=0]   {$q_1$};
    \node[state] (2) [below right of=0]   {$q_2$};
    \node[state] (3) [below right of=1]   {$q_3$};
    \node[state] (4) [right of=3]   {$q_4$};
    \path  (start) edge node {} (0)
    (0) edge node[above] {$\aout[b][c][][n]$} (1)
    (1) edge node[below] {$\ain[c][b][][n]$} (3)
    (0) edge[dashed] node[above] {$\ain[c][b][][n]$} (2)
    (2) edge[dashed] node[below] {$\aout[b][c][][n]$} (3)
    (3) edge node[below] {$\ain$} (4)
    ;
  \end{tikzpicture}
  \quad\qquad
  \aM_{\p[c]} =
  \begin{tikzpicture}[cfsm]
    \node[state] (0) {$q_0$};
    \node[draw=none,fill=none] (start) [left  = 0.3cm  of 0]{};
    \node[state] (1) [below left of=0]   {$q_1$};
    \node[state] (2) [below right of=0]   {$q_2$};
    \node[state] (3) [below right of=1]   {$q_3$};
    \path  (start) edge node {} (0)
    (0) edge node[above] {$\aout[c][b][][n]$} (1)
    (1) edge node[below] {$\ain[b][c][][n]$} (3)
    (0) edge[dashed] node[above] {$\ain[b][c][][n]$} (2)
    (2) edge[dashed] node[below] {$\aout[c][b][][n]$} (3)
    ;
  \end{tikzpicture}
\end{align}
where $\aM_{\p}$ is the CUT.
The split of the mixed choices of $\aM_{\q}$ and $\aM_{\p[c]}$ is
unsafe, because the test including the dashed transitions has a run to
a deadlock configuration despite the fact that $\aM_{\p}$ behaves as
expected.
Note that with insights on the actual notion of well-formedness and of
the projection operation one can deal with mixed choices.
For instance, the well-formedness condition and the projection
operation in~\cite{gt18} yields mixed choice states only when
projecting parallel g-choreographies.
Therefore, it is safe in a mixed-choice state, say $q$, to select a
test starting with one of the output transitions of $q$ and drop all
the others.
Note that this yields \quo{simpler} tests, in line with the principles
of software testing.

Another limitation of the algorithm is its efficiency.
As noted in \cref{sec:generation}, our algorithm is
exponential in the size of the g-choreography.
This is due to the fact that the oracle specification 
used in the algorithm
exhaustively considers all the syntactic sub-trees of the
g-choreographies.
This could be unfeasible for large g-choreographies.
Note however that the oracle specification is a parameter of our
algorithm and, in practice, one can tune it up in order to consider
only \quo{interesting} parts of the g-choreography to target.
Moreover, some optimisations are possible.
A first optimisation can be the reduction of internal choices
generated by the parallel composition as those for $\aG_\mathit{par}$
above.
In fact, those tests are redundant and one would be enough in the
semantics of communicating systems adopted here (where channels are
multisets of messages similar to Erlang's mailboxes).
Note that the tests would not be redundant in the case of
communicating systems interacting through FIFO queues.
Another optimisation relies on the analysis of the syntactic structure
to exclude immaterial sub-trees.
For instance, for the g-choreography
$\gseq[][@][{\gseq[][{\gseq[][{\gint[]}][{\gint[][x][n][y]}]}]}]$ it
is not necessary to check
$\gseq[][{\gseq[][{\gint[]}][{\gint[][x][n][y]}]}]$ because the
sub-tree $\gseq[][{\gint[][x][n][y]}]$ subsumes the runs that \quo{go
  through} the former tree.
A pre-processing of the oracle specification may therefore improve
efficiency.
Note that adopting this approach probably requires a careful
transformation of the oracle specification.
This may not be easy to attain.
Another optimisation comes from the study of some notion of
\quo{dominance} of tests.
The discussion above about mixed-choices is an example: in a
mixed-choice state, the tests with a bias on first-outputs dominate
those starting with inputs.
For instance, the test with solid transitions in \eqref{eq:mixed}
above dominate the one with dashed transitions.

This leads us to consider some other related open questions.
In software testing it is widely accepted that it is unfeasible
to perform a high number of tests.
Hence, test suites are formed by carefully selected tests
that satisfy some \emph{coverage} criteria.
This yields a number of questions that we did not address yet:
What is a good notion of coverage for communicating systems?
Can choreographic models help in identifying good coverage measures?
What heuristics lead to good coverage?
Remarkably, this problem pairs off with the problem of
\emph{concretisation} in model-driven testing~\cite{Pretschner2005}.
Given an abstract test (as the ones we generate), how should it be
concretised to test actual implementations? In fact, the abstract
notion of coverage only considers distributed choices, but actual
implementations may have local branching computations that should also
be covered to some extent.
This probably requires our approach to be combined with existing
approaches to testing.

As said, we took some design decisions to devise our framework.
Alternative approaches are possible.
Firstly, test generation may be done differently when adopting
different types of tests.
In fact, a natural alternative is to take the projection of one
component as the CUT, say $\aM$, and consider as test cases the CFSM
obtained by dualising $\aM$.
Note that this yields a non-local CFSM as a test case; we preferred to
explore first an approach which yields \quo{standard} communicating
systems.

Definition~\ref{def:testgood} formalizes when a test case is
meaningful for a choreography.
It would be also desirable to relate traces of machines that are
test-compliant with the language of the choreography.
Ideally, for a choreography $\aG$ an \emph{adherent} test $\atestcase$
should guarantee that for every $\atestcase$-\emph{compliant} machine
$\aM$ the traces of runs of $\testsys$ that end in a successful
configuration are in $\rlang[\aG]$.
This property cannot be guaranteed by our framework for arbitrary choreographies.
Firstly, the CUT may force causal relations.
For example, consider
$\gint[][a][x][b]; \gint[][b][y][a]; \gint[][a][z][c]$ where \p\ is the CUT.
The event $\aout[b][a][][y]$ should always precede $\aout[a][c][][z]$.
However, this dependency is enforced by \ptp[A] and cannot be checked
by \ptp[B] and \ptp[C] without communication between them.
Secondly, in an asynchronous setting it may be impossible to distinguish some
behaviors of the CUT.
For example, in $\gint[][a][x][b]; \gint[][a][y][b]$
the event $\aout[a][b][][x]$ should always precede $\aout[a][b][][y]$,
but this order is not observable by \ptp[B] in case of asynchronous communication.

In summary, the notion of adherence is not enforceable for all
g-choreographies or all possible implementations of the CUT.  
This hints to the following open problems: the identification of a proper notion of adherence in an asynchronous setting, the identification of
\quo{interesting} subclasses of g-choreographies for which the strict
notion of adherence is meaningful, and the extension of the testing
framework to enforce such notion, either by adding communications
between components or by using non-local machines.

In this work, we consider component testing.
  The level of granularity we adopt implies that participants are components, and our framework is designed to test a single component at a time.
  An intriguing open problem is to apply our framework
  to support integration testing~\cite{jaffar2007testing}.
  In fact, one could think of defining \emph{group} projections, namely
  projection operations that generate communicating systems representing
  the composition of several participants.
  We believe that this approach could pay off when the group onto which
the g-choreography is projected can be partitioned in a set of
\quo{shy} participants that interact only with participants within the
group and others that also interact outside the group.
The former set of participants basically corresponds to units that are
stable parts of the system that and do not need to be (re-)tested
as long as the components in the other group pass some tests.

Instead of concretising abstract tests, one could extract CFSMs from
actual implementations and run the tests on them.
Machines could potentially be extracted
directly from source code.
If however source code was not available
it could still be possible to test components (e.g., by using some
machine learning algorithm to infer the CFSMs from data such as traces).
Note that such technique should be more efficient than
concretisation (because it does not let abstract tests proliferate
into many concrete ones).
Moreover, another advantage of this approach could be that it
enables us to exploit the bottom-up approach of choreographies, where
global views are synthesised from local ones~\cite{lty15}.
The synthesised choreography can be compared with a reference one to
derive tests that are more specific to the implementation at hand.

%%% Local Variables:
%%% mode: latex
%%% TeX-master: "main"
%%% End:

\section{Conclusions \& Related Work}\label{sec:conc}
In software engineering, testing is considered \emph{the}
tool\footnote{
  Regrettably, barred for few exceptions, rigorous formal methods
  that aim to show absence of defects rather than their presence
  are less spread in current practices.
  We cannot embark in a discussion on this state of the matter here.
} for validating software and assuring its quality.
The \emph{Software Engineering Book of Knowledge} available from
\url{http://www.swebok.org} describes \emph{software testing} as (bold
text is ours):
\begin{quote}
  \quo{the dynamic \textbf{verification} of the behaviour of a program
	 on a \textbf{finite} set of \textbf{test cases}, \textbf{suitably
		selected} from the usually \textbf{infinite} executions domain,
	 \textbf{against the expected behavior}.}
\end{quote}
Our framework reflects the description above for model-driven testing
of message-passing systems.
Traditional testing has been classified~\cite{Tretmans99}
according to parameters such as the scale of the system under
test, the source from which tests are derived (e.g., requirements,
models, or code).
There are also classifications according to the specific
characteristics being checked~\cite{oberkampf2010verification}; our work can be
assigned to the category of behavioural testing.

An immediate goal of ours is to experimentally check the suitability
of the test cases obtained with our algorithm.
For this, we plan to identify suitable concretisation mechanisms of the
abstract tests generated by our algorithm, and verify Erlang or Golang
programs.

Since message-passing systems fall under the class of \emph{reactive
  systems} we got inspiration from the work done on model-driven
testing of reactive systems~\cite{bjklp05}.
In particular, we showed that choreographies can, at least to some
extent, be used to automatically generate executable tests and as test
case specifications~\cite{Pretschner2005}.
Technically, we exploited the so-called \emph{projection} operation
of choreographic models.
Here, we gave an abstract notion of projection.
A concrete projection was formalised for the first time
in~\cite{hyc08} (for multiparty session types) and for
g-choreographies in~\cite{gt16,gt17,gt18}, elaborating on the
projection of global graphs~\cite{dy12}.
As discussed in \cref{sec:discussion}, in the future we will also
explore the use of choreographic model-driven testing to address other
problems related to testing message-passing systems.

An interesting theoretical investigation would be to
  explore the relation between our approach and the theory of
  testing~\cite{dh84}.
  At a first glance, our approach corresponds to the must-preorder
  of the testing theory.
  In fact, the notion of test compliance (cf. \cref{def:compliance})
  imposes conditions on all the maximal runs of the CUT in parallel
  with the test.
  However, there are two key differences between the theory of testing
  and our approach which make a precise analysis non trivial.
  The first difference is that we consider asynchronous communications
  and the second is that our tests are \quo{multiparty}, namely tests
  are obtained by composing many CFSMs.
  It might be that the results in~\cite{boreale2002trace}, which
  extend to asynchronous communications the classical theory of
  testing, can be combined with the work in~\cite{dm15} to give a
  suitable theoretical setting to our framework.

According to~\cite{DBLP:books/daglib/0023861}, the generation of test
cases is one of the ways model-based testing can support software
verification.
For example, a component-based testing framework to support online testing
 of choreographed services is proposed in~\cite{Ali2014} for BPMN2 models.
 Among other components, this work sketches a test generation procedure 
 which is however not supported by a formal semantics as we do here.
Our model explicitly features a mechanism for test generation
paired with the notion of an \emph{oracle scheme}
(cf. \cref{def:scheme}) as a precise mechanism to identify
the expected outcome of test cases.
In fact, unlike in most cases, choreographic models contain
enough information about the expected behaviour of the system under
test in order to make accurate predictions.
We believe that this is a highlight of our approach.

%%% Local Variables:
%%% mode: latex
%%% TeX-master: "main"
%%% End:

% \nocite{*}
\bibliographystyle{eptcs}
\bibliography{bib}

\end{document}